\documentclass[12pt]{article}
\usepackage{amsmath}
\usepackage{graphicx,psfrag,epsf}
\usepackage{enumerate}

\newcommand{\blind}{0}
\usepackage{amssymb}
\usepackage{soul}
\usepackage{amsmath}
\usepackage{amsthm}
\usepackage{latexsym}
\usepackage{verbatim}
\usepackage{epsfig}
\usepackage{multirow}
\usepackage{graphicx}
\usepackage{float}
\usepackage{mathrsfs}
\usepackage{mathtools}
			
\usepackage{amsbsy}
\usepackage{amscd}
\usepackage{bm}
\usepackage{graphicx,psfrag,epsf}
\usepackage{enumerate}
\usepackage[authoryear, sort]{natbib}
\setcitestyle{authoryear}
\usepackage{bm}
\usepackage{xcolor}
\definecolor{gmu}{cmyk}{1,0,0.83,0.47}
\usepackage{nicefrac}
\usepackage{url}

\usepackage{caption}
\providecommand{\M}[1]{\mathbf#1}

\DeclareMathOperator{\p}{\mathbf{P}}

\newcommand{\indep}{\rotatebox[origin=c]{90}{$\models$}}

\DeclareMathOperator*{\argmax}{argmax}

\providecommand{\wh}[1]{\widehat{#1}}

\newcommand{\blanco}[1]{  }

\newcommand{\deriv}[3]{%
\ifthenelse{#1 = 1}{\frac{d\,#2}{d\,#3}}{\frac{d^{{#1}} #2}{d{#3}^{{#1}}}}
}

\newcommand{\partials}[3]{%
\ifthenelse{#1 = 1}{\frac{\partial\,#2}{\partial\,#3}}{\frac{\partial^{#1}
    #2}{\partial#3^{#1}}}
}

\newtheorem{lemmachen}{Theorem}

  \newtheorem{lemma}[lemmachen]{Lemma}

\begin{document}

\def\spacingset#1{\renewcommand{\baselinestretch}%
{#1}\small\normalsize} \spacingset{1}

\date{}

\if0\blind
{
  \title{ \large{\bfseries {Relaxing the Assumption of Strongly Non-Informative Linkage Error in Secondary Regression Analysis of Linked Files}}}
   \author{\begin{tabular}{ll} Priyanjali Bukke$^{1}\thanks{Partially supported by NSF grants \#2120318 and \#2411270}$ \thanks{
    Corresponding Author} & $\quad$ Martin Slawski$^{1\star}$ \\
    {\footnotesize \url{fee3qq@virginia.edu}} & $\quad$\;{\footnotesize \url{ebh3ep@virginia.edu}} \end{tabular} \\[3ex]
    \begin{tabular}{l}
    $^{1}${\normalsize Department of Statistics, University of Virginia}\\
    \end{tabular}
   }
  \maketitle

} \fi

\if1\blind
{
  \bigskip
  \bigskip
  \bigskip
  \begin{center}
    {\LARGE\bf Title}
\end{center}
  \medskip
} \fi
\vspace*{-2ex}
\bigskip
\begin{abstract}
Data analysis of files that are a result of linking records from multiple sources are often affected by linkage errors. Records may be linked incorrectly, or their links may be missed. In consequence, it is essential that such errors are taken into account to ensure valid post-linkage inference. Here, we propose an extension to a general framework for regression with linked covariates and responses based on a two-component mixture model, which was developed in prior work. This framework addresses the challenging case of secondary analysis in which only the linked data is available and information about the record linkage process is limited. The extension considered herein relaxes the assumption of {\em strongly non-informative linkage} in the framework according to which linkage does not depend on the covariates used in the analysis, which may be limiting in practice. The effectiveness of the proposed extension is investigated by simulations and a case study.
\end{abstract}

\noindent%

\spacingset{1.45}
\section{Introduction}\label{sec:intro}
\vspace*{-1.5ex}

The term \textit{record linkage} \cite[e.g.,][]{Newcombe, Binette2022, Christen2012} typically is used to refer to the matching of data record-by-record from multiple files and is considered the most granular approach of data integration. Data integration can substantially enhance how we collect and utilize our large existing as well as rapidly growing volume of data. Using record linkage, we can create linked data files richer than individual files in isolation. However, an ongoing challenge of record linkage is that there may be uncertainty when matching records using non-unique or noisy identifiers. The former scenario is commonly a consequence of privacy considerations. For example, HIPAA regulations can restrict personal identifying information to date of birth, zip code, and sex in electronic heath records. In general, the scenarios can also occur, for example, due to variations in formatting, quality (e.g., missingness), or time of collection (e.g., residential addresses), cf.~\cite{Herzog2007}.

Probabilistic record linkage techniques, e.g., those built upon the Fellegi-Sunter model \cite[e.g.,][]{Fellegi69, Binette2022, Herzog2007}, are often used to systematically deal with uncertainty during matching and generate a linked data file by assigning a score to each potential match and then using a threshold to establish matches. At the same time, the choice of the threshold can be challenging, and even with a suitable choice, record linkage is not guaranteed to be error-free: (i) false matches (mismatches) refer to incorrectly matching records belonging to different statistical units and (ii) false non-matches (missed matches) refer to excluding correct matches between records of statistical units. Both types of errors are important, but existing literature tends to focus on the former. This focus is maintained in the work herein. 

In most cases, statistical analyses performed on a linked data file treat all of its linked records at face value. However, missed matches can increase the danger of selection bias similar to non-response in survey data \citep{LittleRubin2019}, and mismatches can cause data contamination that often attenuate estimates of the relationship of interest towards the null \citep[e.g.,][]{Neter65, Scheuren97, Wang2022, Chambers2023}. Alternatively, to avoid the presence of mismatches, analyses may be confined to records considered true matches with near-absolute certainty. In the latter approach the way data is discarded could substantially increase selection bias and decrease statistical power. Hence, the use of suitable methods of post-linkage data analysis (PLDA) is crucial in leveraging the potential of linked data sets.

PLDA can roughly be divided into two settings. I) {\em Primary analysis}: the analyst has the individual files and can jointly conduct record linkage and subsequent data analysis with direct propagation of uncertainty. II) {\em Secondary analysis}: the analyst only has the linked data but not necessarily the individual files. The latter setting is becoming more common and can be especially challenging because information on record linkage may be minimal or absent. 

Earlier work has discussed secondary regression analysis using predictor and response variables from separate files \citep{Chambers2019improved, Lahiri05, Chambers2009}. The approach proposed in \cite{Lahiri05} relies on knowledge or at least estimates of the probability of the correctness 
of each link underlying a (predictor-response)-pair.  The seminal work by \cite{Chambers2009} broadly generalizes this approach and introduces the simplifying ELE (exchangeable linkage error) assumption according to which correct match rates are approximately constant within blocks of observations defined by the same values of categorical variables (e.g., race, state of residence, age group etc.) used during the record linkage process generating the linked file under consideration. A limitation is that these approaches may be inapplicable when the required information on the correctness of links is inaccessible or difficult to estimate, which is rather common in the secondary analysis setting.

In this paper, we focus on a general framework for regression in secondary analysis developed in \cite{slawski2024general} and its precursor \citep{SlawskiDiaoBenDavid2019} that offers flexibility in utilizing information on record linkage of varying degrees to account for potential mismatch error. The framework in \cite{slawski2024general} tends to produce more accurate estimates compared to naive analysis (i.e., not accounting for linkage error) for the regression parameter of primary interest. This flexible framework can accommodate various types of PLDA such as generalized linear models, linear mixed models, Cox regression, and predictive modeling methods \cite[e.g.,][]{cox2025, ben2023novel, FabriziSalvatiSlawski2025}. The framework also exhibits a favorable computational cost (linear in the sample size) and can produce estimates of the mismatch rate and correct match probabilities. 

Of particular interest in this work is the strongly non-informative linkage assumption \cite[cf.~][]{kamat2024analysis} made in that framework. In a nutshell, that assumption states that the variables associated with the record linkage process are statistically independent of the linked data themselves. Although the assumption greatly facilitates inference, it can be limiting in practice as it excludes record linkage scenarios with overlap or dependencies between the matching variables and covariates. For instance, the variable \texttt{age} may be used both to determine eligible links and also be a covariate in the regression analysis to be performed using the linked data. In general, violations of this assumption can lead to significantly biased estimates. Therefore, relying on the assumption may be problematic in practice, and accommodating more complex scenarios is crucial for enabling valid post-linkage data inference in a wide range of scenarios. 

\vspace{1.5ex}

\noindent{}\textbf{Contributions}. In this paper, we show that the strongly non-informative linkage assumption made in the framework in \cite{slawski2024general} can be relaxed. We discuss an extension of the approach in which the underlying record linkage can depend on covariates in the linked data pertaining to the outcome model of interest. In this way, flexibility with accounting for mismatch error can be maintained. Furthermore, the proposed approach is straightforward to implement in standard statistical computing environments such as the software R. The utility of the proposed
formulation is demonstrated via simulations and a case study.

\vspace{1.5ex}

\noindent{}\textbf{Organization}. The remainder of the paper is organized as follows. Section \ref{sec:problem} describes the problem set-up, provides background on the general framework, and presents a simple example setting in which the assumption does not hold. The proposed approach is then discussed in Section \ref{sec:methods}. Sections \ref{sec:sim} and \ref{sec:casestudy} contain an empirical investigation of the performance of the approach. We conclude in Section \ref{sec:conclusion} with a summary of findings and some future directions. 

\vspace{1.5ex}

\noindent{}\textbf{Notation}. Here, we include a summary on notation frequently used in this paper, which follows that in \cite{slawski2024general}. We use the following conventions regarding probability density functions (PDFs): instead of writing 
$f_{\M{x}}(\M{x}_0)$ for the density of a random vector $\M{x}$ evaluated at a point $\M{x}_0$, we drop the symbol in the subscript and simply write $f(\M{x}_0)$ with the convention that the corresponding random variable is inferred from the symbol in the argument. Similar conventions are adopted for joint and conditional PDFs, i.e, we use $f(\M{a}_0, \ldots, \M{z}_0)$ instead of $f_{\M{a} \ldots \M{z}}(\M{a}_0, \ldots, \M{z}_0)$ and $f(\M{x}_0|\M{y}_0)$ instead of $f_{\M{x}|\M{y} = \M{y}_0}(\M{x}_0)$, etc. Note that subscripts in $f$ will be present in case there is no argument. By default, symbols will be boldfaced to indicate vector-valued quantities, with the understanding that boldfaced quantities may also represent scalars as special case; occasionally, normal instead of bold font is used to highlight a scalar quantity. Finally, the dependence of PDFs on parameters is expressed via $f(\cdot\,;\ldots)$, where $\ldots$ represents a list of parameters.  A table summarizing frequently used symbols and notation is given below. \\[3ex]
{\centering
{\small \begin{tabular}{||l|l||l|l|}
\hline
$\mathbb{I}(\cdot)$  & indicator function  & $\M{u} \indep \M{v}$ & random variables $\M{u}$ and $\M{v}$ are independent \\ 
$m$ & mismatch indicator & $f(\M{y}|\M{x}; \bm\theta)$  & conditional PDF of $\M{y}$ given $\M{x}$ (regression setup) \\
$\M{P}(\ldots)$    & probability                    & $\bm\theta$ & parameter describing the $(\M{x},\M{y})$-relationship \\
$\M{E}[\ldots]$    & expectation & $\M{z}$ & covariates informative of mismatch indicator \\      
$[\ldots]^{(t)}$  & iteration counter & $h(\M{z})$ & $\p(m = 0 | \M{z})$\\
$\text{logit}(x)$  & $\log(x/(1-x))$ & $\bm{\gamma}$ & parameter associated with $h$ \\   
    & & $\bm{\theta}^{\ast}, \bm{\gamma}^{\ast}$ \text{etc}. & ``ground truth" parameter values \\ 
\hline    
\end{tabular}}}

\spacingset{1.45}
\section{Problem Setup}\label{sec:problem}
\vspace*{-1.5ex}

This section defines the problem setup and provides background on the general framework for regression involving linked files that is extended herein. We then offer a simple illustrative example in which the assumption of strongly non-informative linkage is violated. 

\subsection{Post-Linkage Data Analysis}

Suppose we have a linked data set $F_{\M{x} \Join \M{y}} = \{ (\M{x}_{i}, \M{y}_{i}) \}_{i = 1}^n$ generated by merging one file containing covariates $F_{\M{x}}^* = \{ \M{x}_j^* \}_{j = 1}^M$ and another file containing responses $F_{\M{y}}^* = \{ \M{y}_k^* \}_{k = 1}^N$. These two individual files include different data pertaining to a common set of statistical units. For simplicity, suppose that every $\M{y}_k^*$, $1 \leq k \leq N$, has one and only one match in $F_{\M{x}}^*$ such that $M \geq N$. Employing variables common to both files to uniquely identify statistical units (``matching variables'') would ideally lead to a new file $F^*_{\M{x} \Join \M{y}} = \{ (\M{x}^*_{\ell_i}, \M{y}^*_{\ell_i}) \}_{i = 1}^\nu$ with $\nu = N$ perfectly matched pairs of observations. However, the actual record linkage instead yields an imperfect new file $F_{\M{x} \Join \M{y}}$ with $n$ matched pairs such that $\M{x}_i \in F_{\M{x}}^*$ and $\M{y}_i \in F_{\M{y}}^*$, for $1 \leq i \leq n \leq N$. This linked data set potentially includes \textit{mismatched pairs} $(\M{x}_i, \M{y}_i) \notin F^*_{\M{x} \Join \M{y}}$ and lacks correct matches $F^*_{\M{x} \Join \M{y}}  \setminus F_{\M{x} \Join \M{y}}$. 

As a special case, for example, if $M = N$ and the $\{x_i^*\}_{i=1}^n$ and $\{y_i^*\}_{i=1}^n$ are permuted against each other, then $(x_i, y_i) = (x_{\tilde{i}}^*, y_{\tilde{j}}^*)$ for some ${\tilde{i}}$ and ${\tilde{j}}$. However, in general, this relationship between the linked data and the variables in the individual files may not be true.
 
During post-linkage data analysis, suppose we want to model response variable(s) $\M{y}$ conditional on covariate(s) $\M{x}$ using parametric regression. Our aim is to conduct inference based on $F_{\M{x} \Join \M{y}}$ that is aligned with inference based on $F^*_{\M{x} \Join \M{y}}$. 

The framework we build upon is motivated by the \textit{secondary analysis} setting, in which the data analyst is given $F_{\M{x} \Join \M{y}}$ but the two individual files $F_{\M{x}}^*$ and $F_{\M{y}}^*$ are unavailable. Missing correct match pairs $F^*_{\M{x} \Join \M{y}}  \setminus F_{\M{x} \Join \M{y}}$ could have occurred but are assumed ignorable in the terminology of missing data. In $F_{\M{x} \Join \M{y}}$, it is unknown which pairs $\{(\M{x}_i, \M{y}_i)\}_{i=1}^n$ are incorrectly matched, but auxiliary information from the underlying record linkage process related to match status of the pairs may be provided (e.g., block indicators\footnote{In record linkage, ``blocking'' refers to the process of considering records as potential matches only if their values agree for a specific subset of the matching variables, termed ``blocking variables''. This procedure partitions the observations into blocks defined by the configurations of these variables and considers links only within blocks.}, scores quantifying the likelihood of a correct match for each pair, the expected overall mismatch error rate in the linked data set, an indicator of pairs that can be treated as correct matches, predictors of match status, etc.).

\subsection{Review of the existing framework}\label{framework}

In an effort to make this paper self-contained, we provide a brief review of the framework detailed in our prior work \citep{slawski2024general}. This framework is based on a two-component mixture model that reflects the distributions contingent on latent mismatch indicators $m_i = \mathbb{I}((\M{x}_i, \M{y}_i) \notin F^*_{\M{x} \Join \M{y}})$, i.e., $m_i = 0$ if $(\M{x}_i, \M{y}_i)$ represents a correct match and $m_i = 1$ otherwise, $1 \leq i \leq n$. In this section, we briefly summarize the framework, which enables adjustment for potential mismatch errors when fitting the regression model of interest using the given linked data.  

\medskip

\noindent{}The existing framework relies on the following assumptions.

\vspace{-0.5ex}
\begin{itemize}
    \item[({\bf{A1}})] \textit{Observation Model and Independence for Mismatches.}
    
    Conditional on $\{ m_i = 1 \}$, $\M{y}_i$ is independent of $\M{x}_i$, whereas if $\{ m_i = 0 \}$, $\M{y}$ follows the true relationship specified by the regression model of interest, $1 \leq i \leq n$. 
    
    \item[({\bf{A2}})] \textit{Latent Mismatch Indicator Model.}
    
    The probability of $\{m_i = 0\}$ can be modeled conditional on variables $\{\M{z}_i\}_{i=1}^n$, related to the underlying record linkage, which are informative of the values of $\{ m_i \}_{i = 1}^n$. Here, $\p(m_i = 0 | \M{z}_i; \boldsymbol{\gamma}) = h(\M{z}_i;\bm{\gamma})$, $1 \leq i \leq n$, for some known function $h$ and unknown parameter $\bm{\gamma}$ (of secondary interest). In the absence of information about match status, the approach is still applicable under the assumption of a constant mismatch rate (i.e., $\p(m_i = 1|\M{z}_i; \boldsymbol{\gamma}) =\alpha$ for all observations $i$, where mismatch rate $\alpha \in (0,1)$) or block-wise constant mismatch rates.
     
    \item[({\bf{A3}})] \textit{Strongly Non-Informative Linkage.}
    
    The $\{ (m_i, \M{z}_i) \}_{i = 1}^n$ are independent of both $\{ \M{x}_i \}_{i = 1}^n$ and $\{ \M{y}_i \}_{i = 1}^n$.
\end{itemize}

  \noindent{}Assumption ($\bf{A1}$) implies in particular that the regression model parameters of interest can be inferred from a subset of correctly matched pairs such that missing the remainder of the underlying correct matches is ignorable, in analogy to missingness at random in a regression setup with missing responses. The assumption of independence between the components of a mismatched pair is satisfied if distinct records are independent as would be the case in a setup in which matching $(\M{x}, \M{y})$ are jointly random, but may be less plausible if the linkage error involves swaps of similar $\M{x}$'s, which is potentially induced via (approximate) agreement of matching variables used during linkage that correlate with $\M{x}$. 
  Violations of this aspect of Assumption ($\bf{A1}$) tend not to have an impact on the estimation of 
  regression parameters \citep{slawski2024general}.
  
The assumption of {\em strongly non-informative linkage} as stated in ($\bf{A3}$) is rather strong, but renders inference more tractable. In particular, it implies that $f(\M{y}|m=0) = f(\M{y}|m=1) = f(\M{y})$. As mentioned in the introduction, the goal of the present paper is to relax this assumption. 

From ($\bf{A1}$) through ($\bf{A3}$), we have that $\M{y}_i$ given $\M{x}_i$ follows the two-component mixture model 
$$\M{y}_i | \M{x}_i \sim f(\M{y}_i | \M{x}_i; \bm\theta)\p(m_i = 0|\M{z}_i; \boldsymbol{\gamma}) + f(\M{y}_i)\p(m_i = 1|\M{z}_i; \boldsymbol{\gamma}), \quad 1 \leq i \leq n.$$ 

\vspace{1.5ex}

    \noindent{}Model fitting and inference can be conducted using (composite) maximum likelihood \citep{Lindsay1988, Varin2011} and the expectation-maximization (EM) algorithm \citep{Dempster1977}. Alternatively, the approach can be cast and fitted in a hierarchal Bayes setup \citep{Gelman2021}. We here focus on the former approach for regression problems. 

    \bigskip
    
\noindent{}\textbf{Estimation}
    
    \noindent{}In this context, the (composite) likelihood resulting from the postulated model with respect to the unknown parameters $(\boldsymbol{\theta}, \boldsymbol{\gamma})$ can be expressed as 
    \begin{equation}\label{eq:clik}
       L(\boldsymbol{\theta}, \boldsymbol{\gamma}) = \prod_{i=1}^n \{ f(\M{y}_i | \M{x}_i; \bm\theta)\p(m_i = 0|\M{z}_i; \boldsymbol{\gamma}) + f(\M{y}_i)\p(m_i = 1|\M{z}_i;\boldsymbol{\gamma}) \}.
    \end{equation}
    The likelihood (\ref{eq:clik}) is ``composite'' as the product over the observation-specific likelihoods may not be a proper likelihood. In other words, it may not coincide with the joint likelihood associated with the entire collection of matched pairs. This is due to the fact that the mismatch indicators of the match pairs may not necessarily be independent. Specifically, linkage of one pair could affect linkage of another (e.g., if one-to-one matching is enforced). Regardless, composite maximum likelihood estimators exhibit favorable properties such as $\sqrt{n}$-consistency and asymptotic normality with an asymptotic covariance matrix which can be expressed in the familiar sandwich form \cite[e.g.,][]{Lindsay1988, Varin2011}). 
    
    \noindent{}To maximize the composite likelihood, we can treat the mismatch indicators $\{m_i\}_{i=1}^n$ as missing data and subsequently utilize the EM-algorithm, iteratively performing two steps as briefly outlined below until convergence \citep{slawski2024general}. 

    \medskip

    \noindent{}{\bf{E}}-step: Given the observed data $\{({\bf{x}}_i, {\bf{y}}_i, {\bf{z}}_i)\}_{i=1}^n$ and estimates of the parameters $(\widehat{\bm\theta}^{(t)}, \widehat{\bm\gamma}^{(t)})$ at any iteration $t$, the expected complete data composite log-likelihood can be computed as below. We recall that $h(\cdot;\bm{\gamma})$ is defined in Assumption ({\bf{A2}}) of the framework stated above.
    \begin{equation}
    \mathscr{L}(\bm\theta, \bm\gamma) =     \sum_{i=1}^n \{\wh{m}_i^{(t)} \log(1-h(\M{z}_i;\bm{\gamma}))) + (1-\wh{m}_i^{(t)})\log(h(\M{z}_i;\bm{\gamma}))\} + \sum_{i=1}^n(1-\wh{m}_i^{(t)})\log(f(\M{y}_i | \M{x}_i; \bm\theta)),
    \end{equation}

    \noindent{}where for $1 \leq i \leq n$,
    \begin{align*}
    \wh{m}_i^{(t)} &= \M{P}(m_i = 1 | ({\bf{x}}_i, {\bf{y}}_i, {\bf{z}}_i); \widehat{\bm\theta}^{(t)}, \widehat{\bm\gamma}^{(t)})  = \dfrac{f(\M{y}_i | \M{x}_i; \widehat{\bm\theta}^{(t)}) \times h(\M{z}_i;\widehat{\bm\gamma}^{(t)})}{f(\M{y}_i | \M{x}_i; \widehat{\bm\theta}^{(t)}) \times h(\M{z}_i;\widehat{\bm\gamma}^{(t)}) +  f(\M{y}_i) \times (1-h(\M{z}_i;\widehat{\bm\gamma}^{(t)}))}. 
    \end{align*}

    \noindent{}{\bf{M}}-step: Given $\mathscr{L}(\bm\theta, \bm\gamma)$, maximization over $\bm \theta$ and $\bm \gamma$ decouples into separate optimization problems:
    \begin{align}\label{eq:EM_plain}
    \begin{split}
    \widehat{\bm\theta}^{(t+1)} &= \argmax_{\bm\theta} \Big\{ \sum_{i=1}^n(1-\wh{m}_i^{(t)})\log(f(\M{y}_i | \M{x}_i; \bm\theta)) \Big\},  \\
    \widehat{\bm\gamma}^{(t+1)} &= \argmax_{\bm\gamma} \Big\{ \sum_{i=1}^n \{\wh{m}_i^{(t)} \log(1-h(\M{z}_i;\bm{\gamma}))) + (1-\wh{m}_i^{(t)})\log(h(\M{z}_i;\bm{\gamma}))\} \Big\}.
    \end{split}
    \end{align}

    \noindent{}So, the update for $\bm\theta$ reduces to the optimization problem for the regression model of interest with additional observation weights. The update for $\bm\gamma$ involves a Bernoulli likelihood parameterized by $\bm\gamma$, which can reduce to a closed form update or fitting a (quasi)-GLM to the $\{\wh{m}_i^{(t)}\}$, depending on the form assumed for the $h$ function.

    \bigskip

    \noindent{}\textbf{Standard Errors and Asymptotic Inference}

    \noindent{}After obtaining the maximizer $(\widehat{\bm\theta}_n, \widehat{\bm\gamma}_n)$ of the composite likelihood $L$ in (\ref{eq:clik}) through the EM-algorithm, estimation of standard errors and asymptotic inference can be conducted based on the properties of composite maximum likelihood estimators. Specifically, we have the following (under suitable regularity conditions). Here, let $(\bm\theta^*, \bm\gamma^*)$ be the population parameters and $\ell = -\log L$:  
    \begin{equation}
    \sqrt{n}\begin{psmallmatrix}
    \widehat{\bm\theta}_n \\
    \widehat{\bm\gamma}_n  
    \end{psmallmatrix}
    \rightarrow N\Big(\begin{psmallmatrix}
    \bm\theta^* \\
    \bm\gamma^*    
    \end{psmallmatrix}, \M{E}[\nabla^2\ell(\bm\theta^*, \bm\gamma^*)]^{-1} \M{E}[\nabla\ell(\bm\theta^*, \bm\gamma^*)\nabla\ell(\bm\theta^*, \bm\gamma^*)^T] \M{E}[\nabla^2\ell(\bm\theta^*, \bm\gamma^*)]^{-1} \Big),
    \end{equation}
    in distribution, where $\nabla\ell$ and $\nabla^2\ell$ denote the gradient and Hessian of $\ell$. The asymptotic covariance matrix can be estimated consistently using the observed data and $(\widehat{\bm\theta}_n, \widehat{\bm\gamma}_n)$.
    
    \medskip
    \vskip7ex
    \noindent{}\textbf{Modeling the Latent Mismatch Indicator}
    
    \noindent{}For the relationship between $\bf{z}$ and $m$, a standard approach is to use a logistic regression model:
  \begin{equation}
   h(\M{z}_i;\bm{\gamma}) = \dfrac{\exp(\gamma_0 + \gamma_1 z_{1,i} + \dots + \gamma_q z_{q,i})}{1+\exp(\gamma_0 + \gamma_1 z_{1,i} + \dots + \gamma_q z_{q,i})} = \dfrac{\exp({\bf{z}}_i^T\bm\gamma)}{1+\exp({\bf{z}}_i^T\bm\gamma))}, \hspace{3ex} 1 \leq i \leq n.
  \end{equation}
  Since the mismatch indicators are not observed, parameter estimation in this model is more challenging than in standard binary regression. Therefore, if available, the underlying proportion of mismatched records in the linked data set can be incorporated by imposing a corresponding constraint: $\Big(\frac{1}{n} \sum_{i=1}^n {\bf{z}}_i\Big)^T (-\boldsymbol{\gamma}) \leq b$, where $b \in \mathbb{R}$ is the logit of the assumed mismatch rate.


   \bigskip

   \noindent{}\textbf{Estimating the Marginal Density}

    \noindent{}The marginal density $f(\M{y})$ is not required to be known and, under assumptions ($\bf{A1}$) and ($\bf{A3}$) above, is not affected by mismatch error. Estimation can mainly be performed either using a \textit{plug-in} or \textit{integrated} approach. In the former approach, $f(\M{y})$ is estimated beforehand based on the $\{\M{y}_i\}_{i=1}^n$ only and substituted as the population quantity. Example options include using kernel density estimation or the empirical probability mass function (if the cardinality of the range of $\M{y}$ is small). Alternatively, we may choose to update $f(\M{y})$ along with parameters $(\bm\theta, \bm\gamma)$ in the EM-iterations. Here, for example, we may leverage the regression model of interest for $\M{y}|\M{x}$ and integrate over $\M{x}$ to successively refine our estimates for $f(\M{y})$.

\subsection{A Motivating Example }

Having provided the necessary context, we are ready to delve into the primary subject of this paper, namely the relaxation of the assumption of strongly-non informative linkage error per ($\bf{A3}$). For illustrative purposes, consider the following setup. We have a linked data set with $n = 1,000$ observations which consists of a single predictor variable $x$ and response variable $y$. The $\{x_i\}_{i=1}^n$ follow a standard Gaussian distribution, and $\{y_i\}_{i=1}^n$ follow a Gaussian distribution with conditional mean ${\bf{E}}[y_i|x_i] = \beta_0^* + \beta_1^*x_i$ and constant variance $\sigma^2$. Here, we set $\boldsymbol{\beta}^* = (\beta_0^*, \beta_1^*)^T = (1, -1)^T$ and $\sigma = 0.25$. Mismatch indicators $\{m_i\}_{i=1}^n$ are produced on the basis of the logistic model below with $\boldsymbol{\gamma}^* = (\gamma_0^*, \gamma_1^*)^T = (2.5, 4.5)^T$. Mismatches are generated by randomly shuffling the subset of the $\{y_i\}_{i=1}^n$ with $m_i = 1$.
\begin{equation}\label{eq:m-model_ex}
\text{logit}({\mathbf{P}\left(m_i=0\middle| x_i\right)})=\gamma_{0}^* +\gamma_{1}^* x_i.  
\end{equation}  

\begin{figure}
\hspace*{-.5ex}\includegraphics[height = 0.31\textwidth]{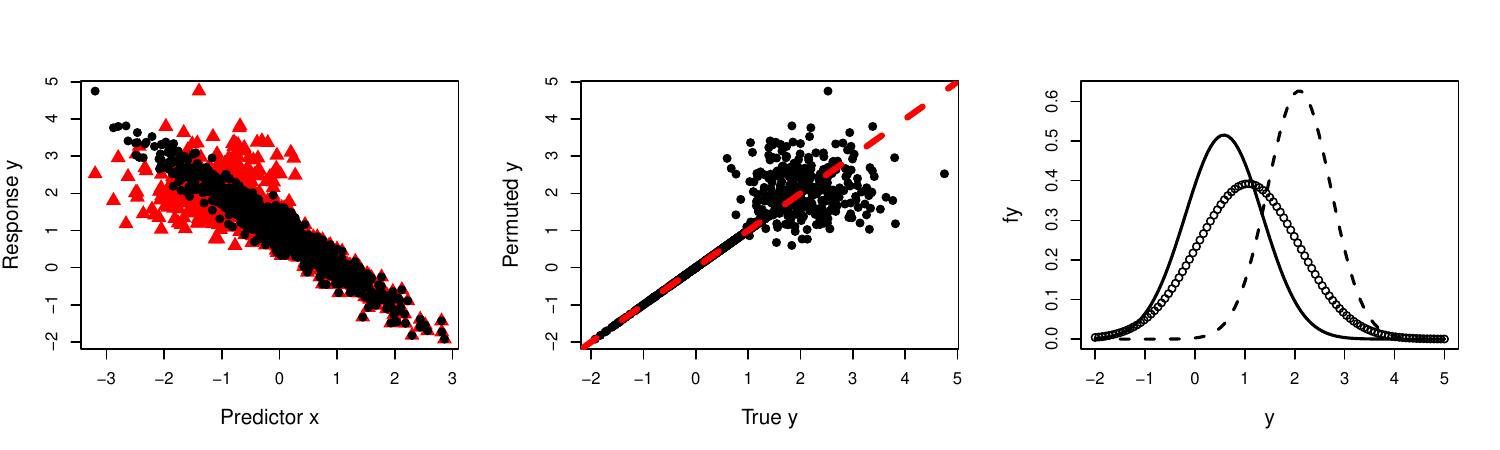}
\hspace*{-1ex}
\vspace*{-7ex}
\caption{{\small Left: scatterplot of true linked data (black) and observed linked data with mismatches (grey) underlaid, Middle: the true $y$ versus the observed permutation of $y$, with the grey line corresponding to the ideal case when there are no mismatches, Right: the marginal density of $y$ estimated for only correct matches (solid), mismatches (dashed), and all the data (dotted).}}\label{fig:intro_ex}
\end{figure}

The linked data is depicted in Figure \ref{fig:intro_ex}. As shown in the first two plots, the linkage error is dependent on the predictor $x$ in the outcome model of interest. Specifically, mismatches with the outcome $y$ tends to occur for lower values of $x$. Consequently, higher $y$ values are more susceptible to mismatch error and we observe that $f(\M{y}|m=0) \neq f(\M{y}|m=1)$ in the third plot. These densities are estimated by assuming Gaussian distributions. 

Suppose we replicate this setting $1,000$ times. In each replication, the observed response is re-sampled and randomly shuffled to induce mismatches between the $\{x_i\}_{i=1}^n$ and $\{y_i\}_{i=1}^n$ in the correctly linked data set. The average mismatch rate over the replications is $\sim 0.30$. Using the linked data set contaminated with mismatches, we then fit a linear regression model either in its standard form (``naive'') or with adjustment based on the framework reviewed in $\S$\ref{framework} (``plain adjustment''). When applying adjustment, the latent mismatch indicator is modeled according to Eq. (\ref{eq:m-model_ex}) despite violating the strongly non-informative linkage assumption.

As presented in Table \ref{tab_intro_results}, the coverage rates for the $\boldsymbol{\beta}^*$ parameter of main interest from both examined approaches show a considerable deviation from the nominal $95\%$ level. While the standard deviations appear comparable, the bias observed likely leads to poorly centered confidence intervals. As a point of reference, we modify the ``plain adjustment" approach based on \eqref{eq:clik} by replacing $f(\M{y})$ with the corrected counterpart $f(\M{y}|m=1)$. The findings suggest that the primary origin of the observed bias is the violation of the strongly non-informative linkage assumption since supplying $f(\M{y}|m=1)$ during adjustment significantly improves both bias and coverage. However, a practical limitation of this solution is that $f(\M{y}|m=1)$ is typically unknown, and unlike $f(\M{y})$, cannot be estimated directly since the $\{ m_i \}_{i = }^n$ are unobserved. The refined adjustment approach presented in the subsequent sections of this paper addresses this limitation. Results derived from this approach are included in Table \ref{tab_intro_results}. This approach extends the framework discussed in Section \ref{framework} by relaxing assumption (\textbf{A3}), enabling estimation of $f(\M{y}|m=1)$ in an integrated fashion, and in turn providing improved performance over the ``plain adjustment". 

\begin{table}[H]
\centering
{\footnotesize
\begin{tabular}{|c|ccc|ccc|ccc|}
\hline
\multirow{2}{*}{} &
  \multicolumn{3}{c|}{$\widehat{\beta}_0$} &
  \multicolumn{3}{c|}{$\widehat{\beta}_1$} &
  \multicolumn{3}{c|}{$\widehat{\sigma}$} \\ \cline{2-10} 
 &
  \multicolumn{1}{c|}{RB} &
  \multicolumn{1}{c|}{SD} &
  CR &
  \multicolumn{1}{c|}{RB} &
  \multicolumn{1}{c|}{SD} &
  CR &
  \multicolumn{1}{c|}{RB} &
  \multicolumn{1}{c|}{SD} &
  CR \\ \hline
Naive &
  \multicolumn{1}{c|}{0.0014} &
  \multicolumn{1}{c|}{0.0079} &
  1 &
  \multicolumn{1}{c|}{0.1187} &
  \multicolumn{1}{c|}{0.012} &
  0 &
  \multicolumn{1}{c|}{1.1964} &
  \multicolumn{1}{c|}{0.0174} &
  0 \\
Adj$^1$ &
  \multicolumn{1}{c|}{0.0293} &
  \multicolumn{1}{c|}{0.0144} &
  0.474 &
  \multicolumn{1}{c|}{-0.0245} &
  \multicolumn{1}{c|}{0.0147} &
  0.681 &
  \multicolumn{1}{c|}{0.0785} &
  \multicolumn{1}{c|}{0.0100} &
  0.540 \\
Adj$^o$ &
  \multicolumn{1}{c|}{0.0004} &
  \multicolumn{1}{c|}{0.0134} &
  0.950 &
  \multicolumn{1}{c|}{-0.0004} &
  \multicolumn{1}{c|}{0.0137} &
  0.957 &
  \multicolumn{1}{c|}{-0.0011} &
  \multicolumn{1}{c|}{0.0073} &
  0.950 \\ \hline
Adj$^*$ &
  \multicolumn{1}{c|}{0.0002} &
  \multicolumn{1}{c|}{0.0101} &
  0.969 &
  \multicolumn{1}{c|}{-0.0002} &
  \multicolumn{1}{c|}{0.0101} &
  0.980 &
  \multicolumn{1}{c|}{-0.0014} &
  \multicolumn{1}{c|}{0.0072} &
  0.949 \\ \hline
\end{tabular}

\vspace{2ex}
\centering
\addtolength{\tabcolsep}{8.5pt}
\begin{tabular}{|c|ccc|ccc|}
\hline
\multirow{2}{*}{} & \multicolumn{3}{c|}{$\widehat{\gamma}_0$}                                            & \multicolumn{3}{c|}{$\widehat{\gamma}_1$}                                            \\ \cline{2-7} 
                  & \multicolumn{1}{c|}{RB}      & \multicolumn{1}{c|}{SD}    & CR    & \multicolumn{1}{c|}{RB}      & \multicolumn{1}{c|}{SD}     & CR    \\ \hline
Adj$^1$              & \multicolumn{1}{c|}{-0.0736} & \multicolumn{1}{c|}{0.1826} & 0.811 & \multicolumn{1}{c|}{-0.5302} & \multicolumn{1}{c|}{0.1726} & 0     \\
Adj$^o$              & \multicolumn{1}{c|}{0.0143}  & \multicolumn{1}{c|}{0.2367} & 0.972 & \multicolumn{1}{c|}{0.0215}  & \multicolumn{1}{c|}{0.5365} & 0.971 \\ \hline
Adj$^*$              & \multicolumn{1}{c|}{0.0157}  & \multicolumn{1}{c|}{0.2623} & 0.906 & \multicolumn{1}{c|}{0.0337}  & \multicolumn{1}{c|}{0.6242} & 0.935 \\ \hline
\end{tabular}}

\caption{{\small Results based on 1k replications. RB -- relative bias; SD -- standard deviation; CG -- coverage rate of $95\%$ confidence intervals. Naive: without adjustment (plain model) ignoring mismatch error; Adj$^1$: with plain adjustment; Adj$^o$: with adjustment and $f(\M{y}|m=1)$ assumed known (``oracle"). Adj$^*$: with adjustment as proposed in this paper.}}\label{tab_intro_results}
\end{table}

\vspace{-5ex}
\spacingset{1.5}
\section{Methods}\label{sec:methods}
\vspace*{-1.5ex}

In extension to the general framework for regression as introduced in Section \ref{framework}, we propose an approach to relax the underlying strongly non-informative linkage assumption and describe the corresponding procedure for model fitting and inference when this assumption may not hold. 

\vspace*{-1.5ex}
\subsection{Relaxing the Strongly Non-Informative Linkage Assumption}

The strongly non-informative linkage assumption, according to which the $\{ (m_i, \M{z}_i) \}_{i = 1}^n$ are independent of both $\{ \M{x}_i \}_{i = 1}^n$ and $\{ \M{y}_i \}_{i = 1}^n$, implies that $f(\M{y}|m=0) = f(\M{y}|m=1) = f(\M{y})$. 
As a result, $f(\M{y}|m=1)$ can be estimated directly from the entire $\{\M{y}_i\}_{i=1}^n$. However, in practice, it is often the case that $f(\M{y}|m=0) \neq f(\M{y}|m=1) \neq f(\M{y})$. To address this limitation, we propose employing a more elaborate scheme for estimating $f(\M{y}|m=1)$ and provide the specific form of the resulting two-component mixture model. 

The following Lemma \ref{lemma1} derives a representation of $f(\M{y}|m=1)$ that is amenable to estimation.

\begin{lemma}\label{lemma1}
Let the probability of a correct match for each observation $(\M{x}_i, \M{y}_i)$ be given by \newline $\p(m_i = 0 | \M{x}_i, \M{y}_i, \M{z}_i; \boldsymbol{\gamma}) = 
\p(m_i = 0 | \M{z}_i; \boldsymbol{\gamma}) = h(\M{z}_i;\bm{\gamma}), \; 1 \leq i \leq n$. Then evaluating $f(\M{y} | m = 1)$ at $\M{y}_i$ can be expressed as 
\begin{align*}
&f(\M{y}_i | m = 1) =  \sum_{j = 1}^n \omega_j(\M{z}_j;\bm{\gamma}) f(\M{y}_i|\M{x}_j;\bm{\theta}), \; 1 \leq i \leq n,\\
&\text{where} \; \; \omega_j(\M{z}_j;\bm{\gamma}) = \dfrac{1 - h(\M{z}_j;\bm{\gamma})}{\sum_{k = 1}^n (1 - h(\M{z}_k;\bm{\gamma}))}, \, 1 \leq j \leq n.
\end{align*} 
%
\end{lemma}
\noindent Under strongly non-informative linkage error, uniform weighting of the summands, i.e., \linebreak $f(\M{y}_i | m_i = 1) = \frac{1}{n} \sum_{j = 1}^n f(\M{y}_i | \M{x}_j; \bm{\theta})$, $1 \leq i \leq n$, is appropriate, but the above representation of $f(\M{y} | m = 1)$ does not rely on that assumption. In particular, the variables in $\M{x}$ and $\M{z}$ may exhibit overlap (with $\M{x} = \M{z}$ being allowed). 

\begin{figure}[H]
  \centering
\includegraphics[height = 0.15\textwidth]{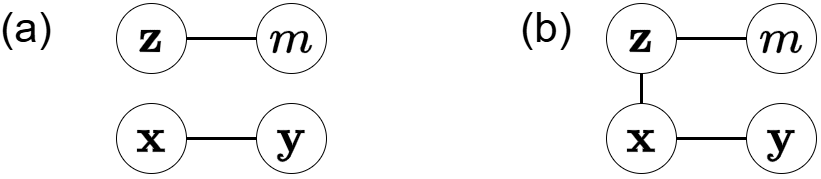}
\vspace*{-2ex}
\caption{{\small Graphical representation of the mixture model and relationships between variables. (a) Framework as in \cite{slawski2024general}. (b) Extended framework developed here with the strongly non-informative linkage assumption relaxed.}}\label{fig:dag}
\end{figure}

Accordingly, we have that $\M{y}_i$ given $\M{x}_i$ follows the two-component mixture model 
\begin{equation}\label{eq:mixture_ext1}
\M{y}_i | \M{x}_i \sim h(\M{z}_i;\bm{\gamma}) f(\M{y}_i | \M{x}_i; \bm\theta) + (1 - h(\M{z}_i;\bm{\gamma})) \sum_{j  = 1}^n \omega_j(\M{z}_j;\bm{\gamma}) f(\M{y}_i | \M{x}_j; \bm\theta), \; 1 \leq i \leq n.
\end{equation}
Figure \ref{fig:dag} summarizes the framework in \cite{slawski2024general} in comparison to the extended framework herein without the strongly informative linkage error assumption. As a remark, we do not consider the situation with a direct edge between $m$ and $\M{y}$ since parameter estimation 
is unlikely to be feasible in this case (cf.~Appendix for more details).   
\noindent In virtue of \eqref{eq:mixture_ext1}, the (composite) likelihood depending on the parameters $(\boldsymbol{\theta}, \boldsymbol{\gamma})$ is given by 
    \begin{equation}\label{eq:clikext}
    \begin{aligned}
       L(\boldsymbol{\theta}, \boldsymbol{\gamma}) &= \prod_{i=1}^n \Big\{ h(\M{z}_i;\bm{\gamma}) f(\M{y}_i | \M{x}_i; \bm\theta) + (1 - h(\M{z}_i;\bm{\gamma})) \sum_{j  = 1}^n \omega_j(\M{z}_j;\bm{\gamma}) f(\M{y}_i | \M{x}_j; \bm\theta) \Big\}, \\
       &= \prod_{i=1}^n \Big\{ \omega_{ii}(\M{z}_i;\bm{\gamma}) f(\M{y}_i|\M{x}_i;\bm{\theta}) + \sum_{j \neq i} \omega_{ij}(\M{z}_i, \M{z}_j;\bm{\gamma}) f(\M{y}_i|\M{x}_j;\bm{\theta}) \Big\}, \\
       &= \prod_{i=1}^n \Big\{ \sum_{j =1}^n \omega_{ij}(\M{z}_i, \M{z}_j;\bm{\gamma}) f(\M{y}_i | \M{x}_j; \bm\theta) \Big\},
    \end{aligned}        
    \end{equation}

\noindent{}where for $1 \leq i \leq n$ and $1 \leq j \leq n$,
\begin{align*}
\omega_{ii}(\M{z}_i;\bm{\gamma}) = h(\M{z}_i;\bm{\gamma}) + \frac{ (1 - h(\M{z}_i;\bm{\gamma}))^2}{\sum_{k = 1}^n (1 - h(\M{z}_k;\bm{\gamma}))}, \quad 
\omega_{ij}(\M{z}_i, \M{z}_j; \bm{\gamma}) = \frac{(1 - h(\M{z}_i;\bm{\gamma}))(1 - h(\M{z}_j;\bm{\gamma}))}{\sum_{k = 1}^n (1 - h(\M{z}_k;\bm{\gamma}))}, \quad j \neq i. 
\end{align*}

\subsection{Estimation and Inference}

We can perform estimation as usual by maximizing the composite likelihood using the EM-algorithm, with the mismatch indicators $\{m_i\}_{i=1}^n$ treated as missing data. Let $\ell_{ij}$ indicate if observation $i$ comes from mixture component $j$ with probability density function $f(\M{y}_i | \M{x}_j; \bm\theta)$. The associated complete data (composite) likelihood is given by 
\begin{equation*}
L^c(\boldsymbol{\theta}, \boldsymbol{\gamma}) = \prod_{i = 1}^n \prod_{j = 1}^n \{ f(\M{y}_i | \M{z}_j; \bm\theta) \omega_{ij}(\M{z}_i, \M{z}_j;\bm{\gamma}) \}^{\ell_{ij}}. 
\end{equation*}

\noindent{}The resulting computations that arise in the E-step and M-step, respectively, are provided below. 

    \medskip

    \noindent{}{\bf{E}}-step: Given the observed data $\{({\bf{x}}_i, {\bf{y}}_i, {\bf{z}}_i)\}_{i=1}^n$ and estimates of the parameters $(\widehat{\bm\theta}^{(t)}, \widehat{\bm\gamma}^{(t)})$ at any iteration $t$, the expected complete data composite log-likelihood can be computed as below. 
    \begin{equation}
    \mathscr{L}(\bm\theta, \bm\gamma) =  \sum_{i = 1}^n \sum_{j = 1}^n \{ \wh{\ell}_{ij}^{(t)} \log{\omega_{ij}(\M{z}_i, \M{z}_j;\bm{\gamma})} \} + \sum_{i = 1}^n \sum_{j = 1}^n \{ \widehat{\ell}_{ij}^{(t)} \log{f(\M{y}_i | \M{x}_j; \bm\theta)} \},
    \end{equation}

    \noindent{}where, for $1 \leq i \leq n$ and $1 \leq j \leq n$,
    \begin{equation*}
    \wh{\ell}_{ij}^{(t)} = \p(\ell_{ij} = 1|({\bf{x}}_j, {\bf{y}}_i, {\bf{z}}_j);\wh{\bm{\theta}}^{(t)}, \wh{\bm{\gamma}}^{(t)}) = \frac{\omega_{ij}(\M{z}_i, \M{z}_j;\wh{\bm{\gamma}}^{(t)}) f(\M{y}_i | \M{x}_j; \wh{\bm{\theta}}^{(t)})}{\sum_{k = 1}^n  \omega_{ik}(\M{z}_i, \M{z}_k;\wh{\bm{\gamma}}^{(t)}) f(\M{y}_i | \M{x}_k; \wh{\bm{\theta}}^{(t)})}.
    \end{equation*}    

    \noindent{}{\bf{M}}-step: Given $\mathscr{L}(\bm\theta, \bm\gamma)$, maximizing over $\bm \theta$ and $\bm \gamma$ decouples into separate optimization problems:
    \begin{align}
    \widehat{\bm\theta}^{(t+1)} &= \argmax_{\bm\theta} \Big\{ \sum_{i = 1}^n \sum_{j = 1}^n \{ \widehat{\ell}_{ij}^{(t)} \log{f(\M{y}_i | \M{x}_j; \bm\theta)} \} \Big\}, \\
    \widehat{\bm\gamma}^{(t+1)} &= \argmax_{\bm\gamma} \Big\{ \sum_{i = 1}^n \sum_{j = 1}^n \{ \wh{\ell}_{ij}^{(t)} \log{\omega_{ij}(\M{z}_i, \M{z}_j;{\bm{\gamma}})} \} \Big\}.
    \end{align}

    \noindent{}Here, the update for $\bm\theta$ reduces to the optimization problem for the regression model of interest with additional observation weights involving $n^2$ data points. The update for $\bm\gamma$ may be performed, for example, using the \texttt{optim()} function in the software $\texttt{R}$. 
 
   Following estimation of the unknown parameters $(\boldsymbol{\theta}, \boldsymbol{\gamma})$, standard errors and asymptotic inference can be conducted as in the framework in prior work (see Section \ref{framework}). 

    \bigskip
    
    \noindent{}We conclude this section by noting that from a computational point view, the price to pay in comparison to the ``plain adjustment" approach based on the framework in $\S$\ref{framework} is an increase from a linear 
    to a quadratic complexity in the number of samples. 
    

\spacingset{1.45}
\section{Simulation Studies}\label{sec:sim}
\vspace*{-1.5ex}

Here, we investigate the performance of the proposed approach through two simulation studies. We first consider a scenario in which there is overlap or correlation between the predictors of the outcome and latent mismatch indicator models. In the second study, the correct match rates are constant within blocks of observations but the blocks are dependent on the predictor in the outcome model. In these studies, empirical run times with the extension are seen to be significantly longer than without (e.g., in the order of up to few minutes vs. always less than a second in R using $1,000$ observations in the first study).

\vspace*{-1.5ex}
\subsection{Overlap or Correlation of the Predictors}
\vspace*{-1.5ex}

We generate a linked data set with $n = 1,000$ observations. The set of predictors are a binary covariate ($d$) that takes the value $0$ or $1$ in equal proportions, a continuous covariate ($x$) that takes values in a linearly spaced grid from $-3$ to $3$, and the associated interaction term ($x \cdot d$). The outcome variable $y$ follows a Gaussian distribution with $\M{E}[y|\M{x}] = \beta_{0}^* + \beta_{1}^* d + \beta_{2}^* x + \beta_{3}^* (x\cdot d)$ and variance $\sigma^2$. The true regression parameters are set to $\boldsymbol{\beta}^* = (\beta_{0}^*, \beta_1^*, \beta_2^*, \beta_{3}^*){^T} = (1,2,-1.5,1)^T$ and $\sigma = 0.25$. Regarding the latent mismatch indicator model, we mimic the scenario when the analyst also has auxiliary data $z_i = \text{logit}(p_i)$, $1 \leq i \leq n$, where the $p_i$ is produced based on the record linkage procedure and represents the probability that $(\M{x}_i, y_i)$ is a correct match. Here, the $\{p_i\}_{i=1}^n$ are drawn i.i.d from a Beta distribution with parameters 4.5 and 0.5. Additionally, variable $d$ may be a predictor such that there is overlap in the outcome and mismatch indicator model. We consider the logistic model below.
\begin{equation}\label{mmod-sim1}
\text{logit}({\mathbf{P}\left(m_i=0\middle| \M{z}_i\right)})=\gamma_{0}^* + \gamma_{1}^* d_i +  \gamma_{2}^* z_i,
\end{equation}
Within this context, we study three settings:
\begin{itemize}
    \item[(i)] The true parameter $\boldsymbol{\gamma^*} = (\gamma_0^*, \gamma_1^*, \gamma_2^*){^T} = (1, -2.5, 1)^T$ and correlation is $\text{Corr}(x, z) = -0.8$. 
    \item[(ii)] Setting (i) modified such that there is no correlation between $x$ and $z$.
    \item[(iii)] Setting (i) modified such that $\gamma_1^* = 0$.
\end{itemize}

Each setting is replicated $1,000$ times. In each replication, the observed response $\{y_i\}_{i=1}^n$ is randomly shuffled to generate mismatches with the predictors $\{\M{x}_i\}_{i=1}^n$. The average mismatch rates in the linked data sets across the replications in each setting (i) -- (iii) are $0.216$, $0.153$, and $0.045$, respectively. Linear regression is then performed (a) ``Naïve'' without adjustment (i.e., ignoring mismatch error), (b) ``Adj$^1$'' with plain adjustment (i.e., based on the framework reviewed in §2.2) assuming an intercept-only $m$-model, (c) ``Adj$^2$'' with plain adjustment assuming $z$ is a predictor in the $m$-model, (d) ``Adj$^3$'' with plain adjustment using both $d$ and $z$ are predictors in the $m$-model (ignoring that this violates assumption {\bf{A2}} in §2.2), (e) Adj$^o$ with plain adjustment but $f(\M{y}|m=1)$ assumed known (``oracle"), and (f) ``Adj$^*$'' with the proposed approach assuming both $d$ and $z$ are $m$-model predictors. Results are reported in Table \ref{tab_sim1}. Estimates for $\boldsymbol{\gamma^}*$ are only provided for ``Adj$^3$", ``Adj$^o$", and ``Adj$^*$" since the other approaches do not consider the full model for correct match status that is stated in Eq.~\eqref{mmod-sim1}. 

\begin{table}[]
\addtolength{\tabcolsep}{-4pt}
\renewcommand{\arraystretch}{0.86}
{\footnotesize \begin{tabular}{|c|c|ccc|ccc|ccc|ccc|}
\hline
\multirow{2}{*}{Setting} &
  \multirow{2}{*}{Method} &
  \multicolumn{3}{c|}{$\widehat{\beta}_0$} &
  \multicolumn{3}{c|}{$\widehat{\beta}_1$ ($d$)} &
  \multicolumn{3}{c|}{$\widehat{\beta}_2$ ($x$)} &
  \multicolumn{3}{c|}{$\widehat{\beta}_3$ ($x \cdot d$)} \\ \cline{3-14} 
 &
   &
  \multicolumn{1}{c|}{Bias} &
  \multicolumn{1}{c|}{SD} &
  CG &
  \multicolumn{1}{c|}{Bias} &
  \multicolumn{1}{c|}{SD} &
  CG &
  \multicolumn{1}{c|}{Bias} &
  \multicolumn{1}{c|}{SD} &
  CG &
  \multicolumn{1}{c|}{Bias} &
  \multicolumn{1}{c|}{SD} &
  CG \\ \hline
\multirow{6}{*}{(i)} &
  Naive &
  \multicolumn{1}{c|}{0.0111} &
  \multicolumn{1}{c|}{0.0245} &
  0.982 &
  \multicolumn{1}{c|}{-0.300} &
  \multicolumn{1}{c|}{0.0491} &
  0 &
  \multicolumn{1}{c|}{0.00971} &
  \multicolumn{1}{c|}{0.0148} &
  0.960 &
  \multicolumn{1}{c|}{0.184} &
  \multicolumn{1}{c|}{0.0302} &
  0 \\ \cline{2-14} 
 &
  Adj$^1$ &
  \multicolumn{1}{c|}{-3.9e-5} &
  \multicolumn{1}{c|}{0.0223} &
  0.954 &
  \multicolumn{1}{c|}{-0.193} &
  \multicolumn{1}{c|}{0.0574} &
  0.020 &
  \multicolumn{1}{c|}{1.32e-4} &
  \multicolumn{1}{c|}{0.0127} &
  0.958 &
  \multicolumn{1}{c|}{0.126} &
  \multicolumn{1}{c|}{0.0350} &
  0.014 \\ \cline{2-14} 
 &
  Adj$^2$ &
  \multicolumn{1}{c|}{-0.00114} &
  \multicolumn{1}{c|}{0.0221} &
  0.956 &
  \multicolumn{1}{c|}{-0.123} &
  \multicolumn{1}{c|}{0.0449} &
  0.252 &
  \multicolumn{1}{c|}{-1.77e-4} &
  \multicolumn{1}{c|}{0.0127} &
  0.956 &
  \multicolumn{1}{c|}{0.0748} &
  \multicolumn{1}{c|}{0.0279} &
  0.0228 \\ \cline{2-14} 
 &
  Adj$^3$ &
  \multicolumn{1}{c|}{6.12e-4} &
  \multicolumn{1}{c|}{0.0222} &
  0.952 &
  \multicolumn{1}{c|}{-0.111} &
  \multicolumn{1}{c|}{0.0422} &
  0.254 &
  \multicolumn{1}{c|}{7.86e-4} &
  \multicolumn{1}{c|}{0.0127} &
  0.946 &
  \multicolumn{1}{c|}{0.0661} &
  \multicolumn{1}{c|}{0.0264} &
  0.268 \\ \cline{2-14} 
 &
   \multicolumn{1}{c|}{Adj$^o$} &
  \multicolumn{1}{c|}{-9.26e-4} &
  \multicolumn{1}{c|}{0.0232} &
  \multicolumn{1}{c|}{0.947} &
  \multicolumn{1}{c|}{0.00101} &
  \multicolumn{1}{c|}{0.0393} &
  \multicolumn{1}{c|}{0.954} &
  \multicolumn{1}{c|}{-1.26e-4} &
  \multicolumn{1}{c|}{0.0132} &
  \multicolumn{1}{c|}{0.95} &
  \multicolumn{1}{c|}{5.09e-4} &
  \multicolumn{1}{c|}{0.0253} &
  \multicolumn{1}{c|}{0.962} \\ \cline{2-14} 
 &
  Adj$^*$ &
  \multicolumn{1}{c|}{-0.00161} &
  \multicolumn{1}{c|}{0.0222} &
  0.950 &
  \multicolumn{1}{c|}{\textbf{0.00217}} &
  \multicolumn{1}{c|}{\textbf{0.0327}} &
  \textbf{0.962} &
  \multicolumn{1}{c|}{-5.09e-4} &
  \multicolumn{1}{c|}{0.0127} &
  0.956 &
  \multicolumn{1}{c|}{\textbf{-9.69e-5}} &
  \multicolumn{1}{c|}{\textbf{0.0193}} &
  \textbf{0.968} \\ \hline
\multirow{6}{*}{(ii)} &
  Naive &
  \multicolumn{1}{c|}{0.0633} &
  \multicolumn{1}{c|}{0.0349} &
  0.737 &
  \multicolumn{1}{c|}{-0.221} &
  \multicolumn{1}{c|}{0.0527} &
  0.015 &
  \multicolumn{1}{c|}{0.0685} &
  \multicolumn{1}{c|}{0.0256} &
  0.179 &
  \multicolumn{1}{c|}{0.0634} &
  \multicolumn{1}{c|}{0.0371} &
  0.522 \\ \cline{2-14} 
 &
  Adj$^1$ &
  \multicolumn{1}{c|}{9.99e-4} &
  \multicolumn{1}{c|}{0.0235} &
  0.957 &
  \multicolumn{1}{c|}{-0.0703} &
  \multicolumn{1}{c|}{0.0396} &
  0.568 &
  \multicolumn{1}{c|}{5.88e-4} &
  \multicolumn{1}{c|}{0.0134} &
  0.955 &
  \multicolumn{1}{c|}{0.0429} &
  \multicolumn{1}{c|}{0.0238} &
  0.530 \\ \cline{2-14} 
 &
  Adj$^2$ &
  \multicolumn{1}{c|}{3.84e-4} &
  \multicolumn{1}{c|}{0.0234} &
  0.960 &
  \multicolumn{1}{c|}{-0.0548} &
  \multicolumn{1}{c|}{0.0379} &
  0.725 &
  \multicolumn{1}{c|}{1.1e-4} &
  \multicolumn{1}{c|}{0.0134} &
  0.961 &
  \multicolumn{1}{c|}{0.0341} &
  \multicolumn{1}{c|}{0.0228} &
  0.664 \\ \cline{2-14} 
 &
  Adj$^3$ &
  \multicolumn{1}{c|}{0.00220} &
  \multicolumn{1}{c|}{0.0234} &
  0.957 &
  \multicolumn{1}{c|}{-0.0432} &
  \multicolumn{1}{c|}{0.0373} &
  0.796 &
  \multicolumn{1}{c|}{0.00152} &
  \multicolumn{1}{c|}{0.0134} &
  0.954 &
  \multicolumn{1}{c|}{0.0233} &
  \multicolumn{1}{c|}{0.0222} &
  0.801 \\ \cline{2-14} 
 &
  \multicolumn{1}{c|}{Adj$^o$} &
  \multicolumn{1}{c|}{0.00134} &
  \multicolumn{1}{c|}{0.0233} &
  \multicolumn{1}{c|}{0.958} &
  \multicolumn{1}{c|}{-0.0014} &
  \multicolumn{1}{c|}{0.0373} &
  \multicolumn{1}{c|}{0.956} &
  \multicolumn{1}{c|}{9.86e-4} &
  \multicolumn{1}{c|}{0.0134} &
  \multicolumn{1}{c|}{0.959} &
  \multicolumn{1}{c|}{-4.85e-4} &
  \multicolumn{1}{c|}{0.0219} &
  \multicolumn{1}{c|}{0.942} \\ \cline{2-14}  
 &
  Adj$^*$ &
  \multicolumn{1}{c|}{0.00133} &
  \multicolumn{1}{c|}{0.0232} &
  0.957 &
  \multicolumn{1}{c|}{\textbf{-0.00115}} &
  \multicolumn{1}{c|}{0.0345} &
  \textbf{0.954} &
  \multicolumn{1}{c|}{9.62e-4} &
  \multicolumn{1}{c|}{0.0132} &
  0.957 &
  \multicolumn{1}{c|}{\textbf{-6.59e-4}} &
  \multicolumn{1}{c|}{0.0202} &
  \textbf{0.950} \\ \hline
\multirow{6}{*}{(iii)} &
  Naive &
  \multicolumn{1}{c|}{0.0115} &
  \multicolumn{1}{c|}{0.0264} &
  0.932 &
  \multicolumn{1}{c|}{-0.0618} &
  \multicolumn{1}{c|}{0.0388} &
  0.632 &
  \multicolumn{1}{c|}{0.00949} &
  \multicolumn{1}{c|}{0.0158} &
  0.895 &
  \multicolumn{1}{c|}{0.0261} &
  \multicolumn{1}{c|}{0.0218} &
  0.771 \\ \cline{2-14} 
 &
  Adj$^1$ &
  \multicolumn{1}{c|}{0.00215} &
  \multicolumn{1}{c|}{0.0243} &
  0.939 &
  \multicolumn{1}{c|}{-0.0231} &
  \multicolumn{1}{c|}{0.0349} &
  0.876 &
  \multicolumn{1}{c|}{7.49e-4} &
  \multicolumn{1}{c|}{0.0141} &
  0.938 &
  \multicolumn{1}{c|}{0.0121} &
  \multicolumn{1}{c|}{0.0193} &
  0.906 \\ \cline{2-14} 
 &
  Adj$^2$ &
  \multicolumn{1}{c|}{0.00229} &
  \multicolumn{1}{c|}{0.0243} &
  0.939 &
  \multicolumn{1}{c|}{-0.0190} &
  \multicolumn{1}{c|}{0.0347} &
  0.894 &
  \multicolumn{1}{c|}{0.00102} &
  \multicolumn{1}{c|}{0.0141} &
  0.939 &
  \multicolumn{1}{c|}{0.00817} &
  \multicolumn{1}{c|}{0.0192} &
  0.932 \\ \cline{2-14} 
 &
  Adj$^3$ &
  \multicolumn{1}{c|}{0.00214} &
  \multicolumn{1}{c|}{0.0243} &
  0.941 &
  \multicolumn{1}{c|}{-0.0190} &
  \multicolumn{1}{c|}{0.0347} &
  0.892 &
  \multicolumn{1}{c|}{9.41e-4} &
  \multicolumn{1}{c|}{0.0141} &
  0.939 &
  \multicolumn{1}{c|}{0.00834} &
  \multicolumn{1}{c|}{0.0192} &
  0.931 \\ \cline{2-14} 
 &
    \multicolumn{1}{c|}{Adj$^o$} &
  \multicolumn{1}{c|}{8.47e-4} &
  \multicolumn{1}{c|}{0.0236} &
  \multicolumn{1}{c|}{0.945} &
  \multicolumn{1}{c|}{-0.00107} &
  \multicolumn{1}{c|}{0.0346} &
  \multicolumn{1}{c|}{0.941} &
  \multicolumn{1}{c|}{1.95e-4} &
  \multicolumn{1}{c|}{0.0137} &
  \multicolumn{1}{c|}{0.948} &
  \multicolumn{1}{c|}{-4.11e-4} &
  \multicolumn{1}{c|}{0.0189} &
  \multicolumn{1}{c|}{0.96} \\ \cline{2-14} 
 &
  Adj$^*$ &
  \multicolumn{1}{c|}{\textbf{9.72e-4}} &
  \multicolumn{1}{c|}{0.0242} &
  0.943 &
  \multicolumn{1}{c|}{\textbf{-5.75e-4}} &
  \multicolumn{1}{c|}{0.0337} &
  \textbf{0.941} &
  \multicolumn{1}{c|}{\textbf{2.22e-4}} &
  \multicolumn{1}{c|}{0.0140} &
  0.941 &
  \multicolumn{1}{c|}{\textbf{-6.71e-4}} &
  \multicolumn{1}{c|}{0.0190} &
  \textbf{0.949} \\ \hline
\end{tabular}}

\vspace{1ex}

\addtolength{\tabcolsep}{-0.05pt}
\renewcommand{\arraystretch}{0.74}
{\footnotesize \begin{tabular}{|c|c|ccc|ccc|ccc|ccc|}
\hline
\multirow{2}{*}{Setting} &
  \multirow{2}{*}{Method} &
  \multicolumn{3}{c|}{$\widehat{\sigma}$} &
  \multicolumn{3}{c|}{$\widehat{\gamma}_0$} &
  \multicolumn{3}{c|}{$\widehat{\gamma}_1$ ($d$)} &
  \multicolumn{3}{c|}{$\widehat{\gamma}_2$ ($z$)} \\ \cline{3-14} 
 &
   &
  \multicolumn{1}{c|}{Bias} &
  \multicolumn{1}{c|}{SD} &
  CG &
  \multicolumn{1}{c|}{Bias} &
  \multicolumn{1}{c|}{SD} &
  CG &
  \multicolumn{1}{c|}{Bias} &
  \multicolumn{1}{c|}{SD} &
  CG &
  \multicolumn{1}{c|}{Bias} &
  \multicolumn{1}{c|}{SD} &
  CG \\ \hline
\multirow{6}{*}{(i)} &
  Naive &
  \multicolumn{1}{c|}{0.118} &
  \multicolumn{1}{c|}{0.0188} &
  0 &
  \multicolumn{1}{c|}{} &
  \multicolumn{1}{c|}{} &
   &
  \multicolumn{1}{c|}{} &
  \multicolumn{1}{c|}{} &
   &
  \multicolumn{1}{c|}{} &
  \multicolumn{1}{c|}{} &
   \\ \cline{2-14} 
 &
  Adj$^1$ &
  \multicolumn{1}{c|}{0.0576} &
  \multicolumn{1}{c|}{0.0163} &
  0.004 &
  \multicolumn{1}{c|}{} &
  \multicolumn{1}{c|}{} &
   &
  \multicolumn{1}{c|}{} &
  \multicolumn{1}{c|}{} &
   &
  \multicolumn{1}{c|}{} &
  \multicolumn{1}{c|}{} &
   \\ \cline{2-14} 
 &
  Adj$^2$ &
  \multicolumn{1}{c|}{0.0287} &
  \multicolumn{1}{c|}{0.0110} &
  0.450 &
  \multicolumn{1}{c|}{} &
  \multicolumn{1}{c|}{} &
   &
  \multicolumn{1}{c|}{} &
  \multicolumn{1}{c|}{} &
   &
  \multicolumn{1}{c|}{} &
  \multicolumn{1}{c|}{} &
   \\ \cline{2-14} 
 &
  Adj$^3$ &
  \multicolumn{1}{c|}{0.0230} &
  \multicolumn{1}{c|}{0.00974} &
  0.298 &
  \multicolumn{1}{c|}{3.124} &
  \multicolumn{1}{c|}{4.302} &
  0.594 &
  \multicolumn{1}{c|}{-1.183} &
  \multicolumn{1}{c|}{4.230} &
  0.80 &
  \multicolumn{1}{c|}{-0.383} &
  \multicolumn{1}{c|}{0.165} &
  0.37 \\ \cline{2-14} 
 &
  Adj$^o$ &
  \multicolumn{1}{c|}{-4.36e-4} &
  \multicolumn{1}{c|}{0.00713} &
  0.947 &
  \multicolumn{1}{c|}{0.863} &
  \multicolumn{1}{c|}{3.121} &
  0.897 &
  \multicolumn{1}{c|}{-0.914} &
  \multicolumn{1}{c|}{3.074} &
  0.91 &
  \multicolumn{1}{c|}{0.0327} &
  \multicolumn{1}{c|}{0.213} &
  0.953 \\ \cline{2-14} 
 &
  Adj$^*$ &
  \multicolumn{1}{c|}{\textbf{-2.95e-4}} &
  \multicolumn{1}{c|}{\textbf{0.00693}} &
  \textbf{0.954} &
  \multicolumn{1}{c|}{\textbf{0.452}} &
  \multicolumn{1}{c|}{\textbf{1.610}} &
  \textbf{0.962} &
  \multicolumn{1}{c|}{\textbf{-0.482}} &
  \multicolumn{1}{c|}{\textbf{1.550}} &
  \textbf{0.95} &
  \multicolumn{1}{c|}{\textbf{0.0210}} &
  \multicolumn{1}{c|}{0.224} &
  \textbf{0.95} \\ \hline
\multirow{7}{*}{(ii)} &
  Naive &
  \multicolumn{1}{c|}{0.215} &
  \multicolumn{1}{c|}{0.0364} &
  0 &
  \multicolumn{1}{c|}{} &
  \multicolumn{1}{c|}{} &
   &
  \multicolumn{1}{c|}{} &
  \multicolumn{1}{c|}{} &
   &
  \multicolumn{1}{c|}{} &
  \multicolumn{1}{c|}{} &
   \\ \cline{2-14} 
 &
  Adj$^1$ &
  \multicolumn{1}{c|}{0.0171} &
  \multicolumn{1}{c|}{0.00914} &
  0.5 &
  \multicolumn{1}{c|}{} &
  \multicolumn{1}{c|}{} &
   &
  \multicolumn{1}{c|}{} &
  \multicolumn{1}{c|}{} &
   &
  \multicolumn{1}{c|}{} &
  \multicolumn{1}{c|}{} &
   \\ \cline{2-14} 
 &
  Adj$^2$ &
  \multicolumn{1}{c|}{0.0123} &
  \multicolumn{1}{c|}{0.00805} &
  0.743 &
  \multicolumn{1}{c|}{} &
  \multicolumn{1}{c|}{} &
   &
  \multicolumn{1}{c|}{} &
  \multicolumn{1}{c|}{} &
   &
  \multicolumn{1}{c|}{} &
  \multicolumn{1}{c|}{} &
   \\ \cline{2-14} 
 &
  Adj$^3$ &
  \multicolumn{1}{c|}{0.00662} &
  \multicolumn{1}{c|}{0.00736} &
  0.862 &
  \multicolumn{1}{c|}{0.140} &
  \multicolumn{1}{c|}{0.377} &
  0.944 &
  \multicolumn{1}{c|}{0.691} &
  \multicolumn{1}{c|}{0.313} &
  0.495 &
  \multicolumn{1}{c|}{-0.0440} &
  \multicolumn{1}{c|}{0.148} &
  0.916 \\ \cline{2-14} 
 &
   Adj$^o$ &
  \multicolumn{1}{c|}{-7.75e-4} &
  \multicolumn{1}{c|}{0.00675} &
  0.948 &
  \multicolumn{1}{c|}{-0.00355} &
  \multicolumn{1}{c|}{0.374} &
  0.937 &
  \multicolumn{1}{c|}{-0.0636} &
  \multicolumn{1}{c|}{0.349} &
  0.979 &
  \multicolumn{1}{c|}{0.0306} &
  \multicolumn{1}{c|}{0.166} &
  0.948 \\ \cline{2-14} 
 &
  Adj$^*$ &
  \multicolumn{1}{c|}{\textbf{-7.28e-4}} &
  \multicolumn{1}{c|}{\textbf{0.00668}} &
  \textbf{0.954} &
  \multicolumn{1}{c|}{\textbf{-0.00446}} &
  \multicolumn{1}{c|}{0.376} &
  0.918 &
  \multicolumn{1}{c|}{\textbf{-0.0595}} &
  \multicolumn{1}{c|}{0.396} &
  \textbf{0.940} &
  \multicolumn{1}{c|}{\textbf{0.0291}} &
  \multicolumn{1}{c|}{0.166} &
  \textbf{0.954} \\ \cline{2-14} 
 &
  Naive &
  \multicolumn{1}{c|}{0.0480} &
  \multicolumn{1}{c|}{0.0209} &
  0.00276 &
  \multicolumn{1}{c|}{} &
  \multicolumn{1}{c|}{} &
   &
  \multicolumn{1}{c|}{} &
  \multicolumn{1}{c|}{} &
   &
  \multicolumn{1}{c|}{} &
  \multicolumn{1}{c|}{} &
   \\ \hline
\multirow{5}{*}{(iii)} &
  Adj$^1$ &
  \multicolumn{1}{c|}{0.00691} &
  \multicolumn{1}{c|}{0.00705} &
  0.821 &
  \multicolumn{1}{c|}{} &
  \multicolumn{1}{c|}{} &
   &
  \multicolumn{1}{c|}{} &
  \multicolumn{1}{c|}{} &
   &
  \multicolumn{1}{c|}{} &
  \multicolumn{1}{c|}{} &
   \\ \cline{2-14} 
 &
  Adj$^2$ &
  \multicolumn{1}{c|}{0.00449} &
  \multicolumn{1}{c|}{0.00674} &
  0.917 &
  \multicolumn{1}{c|}{} &
  \multicolumn{1}{c|}{} &
   &
  \multicolumn{1}{c|}{} &
  \multicolumn{1}{c|}{} &
   &
  \multicolumn{1}{c|}{} &
  \multicolumn{1}{c|}{} &
   \\ \cline{2-14} 
 &
  Adj$^3$ &
  \multicolumn{1}{c|}{0.00454} &
  \multicolumn{1}{c|}{0.00680} &
  0.891 &
  \multicolumn{1}{c|}{1.813} &
  \multicolumn{1}{c|}{3.418} &
  0.846 &
  \multicolumn{1}{c|}{-0.592} &
  \multicolumn{1}{c|}{3.352} &
  0.868 &
  \multicolumn{1}{c|}{-0.148} &
  \multicolumn{1}{c|}{0.349} &
  0.854 \\ \cline{2-14} 
 &
  Adj$^o$ &
  \multicolumn{1}{c|}{-8.03e-4} &
  \multicolumn{1}{c|}{0.00643} &
  0.934 &
  \multicolumn{1}{c|}{0.751} &
  \multicolumn{1}{c|}{3.176} &
  0.903 &
  \multicolumn{1}{c|}{-0.755} &
  \multicolumn{1}{c|}{3.055} &
  0.911 &
  \multicolumn{1}{c|}{0.0507} &
  \multicolumn{1}{c|}{0.325} &
  0.944 \\ \cline{2-14} 
 &
  Adj$^*$ &
  \multicolumn{1}{c|}{\textbf{-8.23e-4}} &
  \multicolumn{1}{c|}{0.00649} &
  \textbf{0.935} &
  \multicolumn{1}{c|}{\textbf{0.484}} &
  \multicolumn{1}{c|}{\textbf{2.022}} &
  \textbf{0.964} &
  \multicolumn{1}{c|}{\textbf{-0.488}} &
  \multicolumn{1}{c|}{\textbf{1.878}} &
  \textbf{0.966} &
  \multicolumn{1}{c|}{\textbf{0.0329}} &
  \multicolumn{1}{c|}{0.341} &
  \textbf{0.956} \\ \hline
\end{tabular}}

\caption{\small Results based on 1k replications. SD -- standard deviation; CG -- coverage rate of $95\%$ confidence intervals. Results improved by the proposed approach relative to the other approaches (excluding Adj$^o$) are indicated in bold. Settings: (i) overlap and correlation, (ii) overlap, (iii) correlation.}\label{tab_sim1}
\end{table}

\noindent As expected, adjustment methods consistently improve estimation of $\beta^*$ for the outcome model of interest relative to performing naïve analysis. Throughout the settings (i)--(iii),
all methods exhibit negligible bias for $\beta_2^*$ (corresponding to $x)$ and adequate coverage levels. However, for the coefficients $\beta_1^*$ and $\beta_3^*$ (corresponding to $d$ and $x \cdot d$, respectively), only the proposed approach achieves negligible bias and adequate coverage. The bias 
and undercoverage of the other adjustment methods exacerbates as the entanglement with the $m$-model increases, with 
setting (iii) and setting (i) corresponding to the lowest and highest levels of entanglement, respectively. The bias and undercoverage of the other adjustment methods also worsen with the degree
of misspecification regarding the $m$-model: approach ``$\text{Adj}^{1}$", which assumes an intercept-only model, exhibits roughly twice the bias of ``$\text{Adj}^{3}$", which works with the correct model, but ignores the lack of strong non-informativeness. Even though the proposed approach is arguably
more complex than the ``plain adjustment" methods, its standard deviations are comparable or even lower. The picture is overall similar for $\sigma$ and the coefficients $\gamma_0^*, \gamma_1^*, \gamma_2^*$ associated with the $m$-model.  It is seen that the performance of the proposed approach is rather close to that of the oracle adjustment.


\vspace*{-1.5ex}
\subsection{Exchangeable Linkage Errors}
\vspace*{-1.5ex}

We consider a linked data set of $n = 1,000$ observations containing $\{x_i\}_{i=1}^n$ and $\{y_i\}_{i=1}^n$ similar to §2.3. However, in this study, mismatches are generated according to the exchangeable linkage error mechanism \citep{Chambers2009}. The $\{x_i\}_{i=1}^n$ are standard normal and $\{y_i\}_{i=1}^n$ follow a Gaussian distribution with conditional mean $\M{E}[y|\M{x}] = \beta_{0}^* + \beta_{1}^* x$ and constant variance $\sigma^2$. The true regression parameters are set to $\boldsymbol{\beta}^* = (\beta_{0}^*, \beta_1^*){^T} = (1,-1)^T$ and $\sigma = 0.25$. The probability of a correct match is constant within blocks of the observations that are defined by the variable $\{z_i\}_{i=1}^n$. There are two blocks of roughly equal size (i.e., $\sim 500$) such that, for a given observation, $z_i = 1$ when $x_i \leq 0$ and $z_i = 2$ when $x_i > 0$. The correct match probability model is as below, with sum-to-zero contrasts for the block effects and the parameters $\boldsymbol{\gamma}^* = (\gamma_{1}^*, \gamma_{2}^*) \approx (-1.19, -2.15)$ set based on true overall correct match probabilities of about $0.28$ and $0.97$ within the blocks, respectively.    
\begin{equation}\label{mmod-sim2}
\text{logit}({\mathbf{P}\left(m_i=0\middle| \M{z}_i\right)})=\gamma_{0}^* + \gamma_{1}^* \mathbb{I}_{(z_i = 1)} +  \gamma_{2}^* \mathbb{I}_{(z_i = 2)}, \text{ subject to } \gamma_{1}^* + \gamma_{2}^* = 0,
\end{equation}

This setting is replicated $1,000$ times. In each replication, the observed responses $\{y_i\}_{i=1}^n$ are permuted randomly within their corresponding block to generate mismatches with the predictors $\{\M{x}_i\}_{i=1}^n$. Linear regression is then performed (a) ``Naïve'' without adjustment (i.e., ignoring mismatch error), (b) using a bias-corrected estimating equation assuming the correct match probabilities are known, (c) with adjustment as in the framework in §2.2, (d) Adj$^o$ with adjustment but $f(\M{y}|m=1)$ assumed known (``oracle"), and (e) ``Adj$^*$'' with the proposed approach assuming the block is a predictor in the $m$-model. Regarding (ii), we include varied types of correction discussed by \cite{Chambers2009}: Lahiri-Larsen (``ELE$^a$''), Kovacevic (``ELE$^b$''), BLUE (``ELE$^c$''), and ratio-type (``ELE$^r$''). In terms of (c), we run adjustment either assuming an intercept-only $m$-model (``Adj$^1$'') or assuming that the block variable is a predictor in the $m$-model (``Adj$^2$''). Results are reported in Table \ref{tab_sim2}. Estimates for $\boldsymbol{\gamma^}*$ are only provided for Adj$^2$ and Adj$^*$ since the other approaches do not consider the full model stated in Eq.~\eqref{mmod-sim2}.

While the naive approach estimates $\beta_0$ relatively well, it demonstrates the poorest performance in estimating $\beta_1$. The ELE methods typically yield minimum bias and standard deviation but often result in over-coverage. The proposed approach performs similarly to the plain adjustment method, with its overall advantage primarily being improved coverage for $\beta_1$. Comparing ``Adj$^2$" and ``Adj$^*$", we find that both lead to under-coverage for $\boldsymbol{\gamma^}*$. Adj$^2$ tends to have better results for $\boldsymbol{\gamma^}*$, but ``Adj$^*$" tends to improve estimation for $\boldsymbol{\beta^}*$, including better coverage and a significantly lower standard deviation. Regarding the latter, we note that the efficiency of ``Adj$^*$" is essentially as good as the most efficient among the four of the ELE methods (ELE$^{a}$), without requiring the block-wise mismatch rates to be known. 

\begin{table}[]
{\footnotesize \begin{tabular}{|c|ccc|ccc|ccc|ccc|}
\hline
 &
  \multicolumn{3}{c|}{$\widehat{\beta}_0$} &
  \multicolumn{3}{c|}{$\widehat{\beta}_1$} &
  \multicolumn{3}{c|}{$\widehat{\gamma}_0$} &
  \multicolumn{3}{c|}{$\widehat{\gamma}_1$} \\ \hline
 &
  \multicolumn{1}{c|}{RB} &
  \multicolumn{1}{c|}{SD} &
  CG &
  \multicolumn{1}{c|}{RB} &
  \multicolumn{1}{c|}{SD} &
  CG &
  \multicolumn{1}{c|}{RB} &
  \multicolumn{1}{c|}{SD} &
  CG &
  \multicolumn{1}{c|}{RB} &
  \multicolumn{1}{c|}{SD} &
  CG \\ \hline
Naive &
  \multicolumn{1}{c|}{-1.31e-4} &
  \multicolumn{1}{c|}{0.0087} &
  1 &
  \multicolumn{1}{c|}{0.138} &
  \multicolumn{1}{c|}{0.016} &
  0 &
  \multicolumn{1}{c|}{} &
  \multicolumn{1}{c|}{} &
   &
  \multicolumn{1}{c|}{} &
  \multicolumn{1}{c|}{} &
   \\ \hline
ELE$^a$ &
  \multicolumn{1}{c|}{-2.47e-4} &
  \multicolumn{1}{c|}{0.00786} &
  1 &
  \multicolumn{1}{c|}{4.24e-4} &
  \multicolumn{1}{c|}{0.00993} &
  0.999 &
  \multicolumn{1}{c|}{} &
  \multicolumn{1}{c|}{} &
   &
  \multicolumn{1}{c|}{} &
  \multicolumn{1}{c|}{} &
   \\ \hline
ELE$^b$ &
  \multicolumn{1}{c|}{-2.63e-4} &
  \multicolumn{1}{c|}{0.00795} &
  1 &
  \multicolumn{1}{c|}{0.0016} &
  \multicolumn{1}{c|}{0.0388} &
  0.977 &
  \multicolumn{1}{c|}{} &
  \multicolumn{1}{c|}{} &
   &
  \multicolumn{1}{c|}{} &
  \multicolumn{1}{c|}{} &
   \\ \hline
ELE$^c$ &
  \multicolumn{1}{c|}{2e-4} &
  \multicolumn{1}{c|}{0.00997} &
  0.991 &
  \multicolumn{1}{c|}{7.45e-5} &
  \multicolumn{1}{c|}{0.0114} &
  0.989 &
  \multicolumn{1}{l|}{} &
  \multicolumn{1}{l|}{} &
   &
  \multicolumn{1}{l|}{} &
  \multicolumn{1}{l|}{} &
   \\ \hline
ELE$^r$ &
  \multicolumn{1}{c|}{-2.5e-4} &
  \multicolumn{1}{c|}{0.00787} &
  1 &
  \multicolumn{1}{c|}{-0.0012} &
  \multicolumn{1}{c|}{0.015} &
  0.996 &
  \multicolumn{1}{l|}{} &
  \multicolumn{1}{l|}{} &
   &
  \multicolumn{1}{l|}{} &
  \multicolumn{1}{l|}{} &
   \\ \hline
Adj$^1$ &
  \multicolumn{1}{c|}{7.47e-5} &
  \multicolumn{1}{c|}{0.015} &
  0.962 &
  \multicolumn{1}{c|}{0.0182} &
  \multicolumn{1}{c|}{0.0155} &
  0.832 &
  \multicolumn{1}{l|}{} &
  \multicolumn{1}{l|}{} &
   &
  \multicolumn{1}{l|}{} &
  \multicolumn{1}{l|}{} &
   \\ \hline
Adj$^2$ &
  \multicolumn{1}{c|}{0.0038} &
  \multicolumn{1}{c|}{0.0139} &
  0.948 &
  \multicolumn{1}{c|}{0.005} &
  \multicolumn{1}{c|}{0.0139} &
  0.932 &
  \multicolumn{1}{c|}{0.998} &
  \multicolumn{1}{c|}{1.101} &
  0.020 &
  \multicolumn{1}{c|}{-0.109} &
  \multicolumn{1}{c|}{1.093} &
  0.893 \\ \hline
  Adj$^o$ &
  \multicolumn{1}{c|}{-0.00454} &
  \multicolumn{1}{c|}{0.014} &
  0.948 &
  \multicolumn{1}{c|}{0.00571} &
  \multicolumn{1}{c|}{0.0139} &
  0.938 &
  \multicolumn{1}{c|}{1.012} &
  \multicolumn{1}{c|}{1.755} &
  0.246 &
  \multicolumn{1}{c|}{-0.521} &
  \multicolumn{1}{c|}{1.743} &
  0.44 \\ \hline
Adj$^*$ &
  \multicolumn{1}{c|}{-0.0031} &
  \multicolumn{1}{c|}{0.0100} &
  0.967 &
  \multicolumn{1}{c|}{0.0044} &
  \multicolumn{1}{c|}{0.0105} &
  \textbf{0.957} &
  \multicolumn{1}{c|}{1.064} &
  \multicolumn{1}{c|}{1.14} &
  0.221 &
  \multicolumn{1}{c|}{-0.579} &
  \multicolumn{1}{c|}{1.145} &
  0.216 \\ \hline
\end{tabular}}
\caption{\small Results for the second simulation study based on 1k replications. RB -- relative bias; SD -- standard deviation; CG -- coverage rate of $95\%$ confidence intervals. Results improved by the proposed approach relative to the other approaches (excluding Adj$^o$) are indicated in bold.}\label{tab_sim2}
\end{table}

\spacingset{1.45}
\section{Case Study}\label{sec:casestudy}
\vspace*{-1.5ex}

We consider the Italian Survey on Household Income and Wealth (SHIW) data discussed in \cite{Wang2022} and \cite{Tancredi15}, which is collected every two years [\cite{shiw}]. The relationship of interest is between individual net disposable income in 2008 (predictor ``$x$'') and 2010 (response ``$y$''), which are variables available in separate files. Records for the same individual across both files can be linked using their household identifier in combination with their household member identifier. These matching variables are present in both files and expected to uniquely identify individuals. In consequence, the record linkage is expected to produce a linked data set free of mismatch errors. 

\begin{figure}[H]
\centering
\hspace*{-.5ex}\includegraphics[height = 0.50\textwidth]{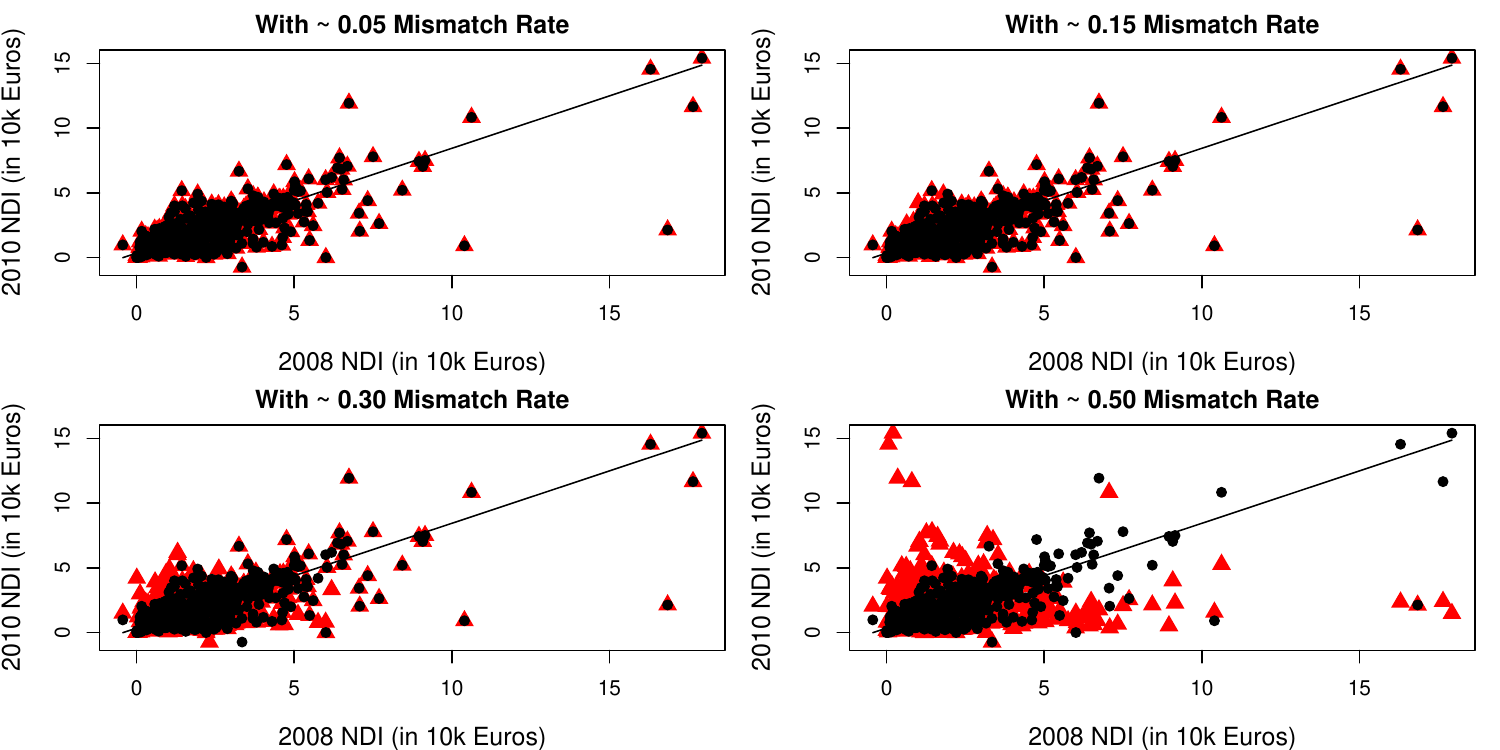}
\hspace*{-1ex}
\vspace*{-2ex}
\caption{Linked SHIW data with the net disposable income (in 10k Euros) from 2010 regressed on that of 2008. Scatterplots of the linked data including mismatches (grey) overlaid, for reference, by the original linked data without the mismatch error (black) along with the estimated robust regression line based on this original data (``oracle'').}\label{fig:cs_est1}
\end{figure}

We here introduce mismatches between the predictor and response to investigate the performance of the proposed approach on real data. For the purpose of this illustration, we utilize a simple random sample of $1k$ observations from the linked data set. We permute subsets of the linked data with varying sizes (approximately $5\%$, $15\%$, $30\%$, and $50\%$) based on the following logistic model for the mismatch indicator. The true parameter values $\boldsymbol{\gamma}^* = (\gamma_0^*, \gamma_1^*)$ for each mismatch rate setting are reported in Table \ref{tab_cs}. Here, we consider a scenario in which mismatches are more likely for observations with atypical or potentially under-represented predictor values. The original linked data along with the varying versions with mismatches are shown in Figure \ref{fig:cs_est1}.

\vspace{-3ex}
\begin{equation}    
\text{logit}({\mathbf{P}\left(m_i=0\middle| x_i\right)})=\gamma_{0} + \gamma_{1} \hspace{1ex} (\text{median}(\{x_i\}_{i=1}^n) - x_i)^2.
\end{equation}
\vspace{-2.5ex}

\vspace{-2ex}

Based on visual exploration of the true linked data, we assume a Gaussian regression model for the predictor-response relationship. We consider five analyses using the linked data sets with mismatches generated: (1) ``Naive'': without adjustment ignoring mismatch error in the data, (2) Adj$^1$: with ``plain'' adjustment using the framework in $\S$\ref{framework} and an intercept-only model for the mismatch indicator, (3) Adj$^{1c}$: same as Adj$^1$ but with upper bounds on the mismatch rate corresponding to the setting, (4) Adj$^2$: with ``plain'' adjustment using the framework in $\S$\ref{framework} and the correct model for the mismatch indicator, and (5) Adj$^*$: with adjustment using the proposed approach. As a reference, we use an ``Oracle'' approach in which we perform analysis on the original linked data free of the mismatch errors. In the ``Naive'' and ``Oracle'' approaches, robust regression is performed using the \texttt{rlm()} function in the \texttt{MASS} package [\cite{ripley2013package}].  Here, robust regression is fitted to mitigate the impact of outliers observed during the analysis. The mixture model in the adjustment approaches can estimate the regression parameters of interest without high sensitivity to outliers \citep{SlawskiDiaoBenDavid2019, FabriziSalvatiSlawski2025}. 

\begin{figure}[H]
\centering
\hspace*{-.5ex}\includegraphics[height = 0.51\textwidth]{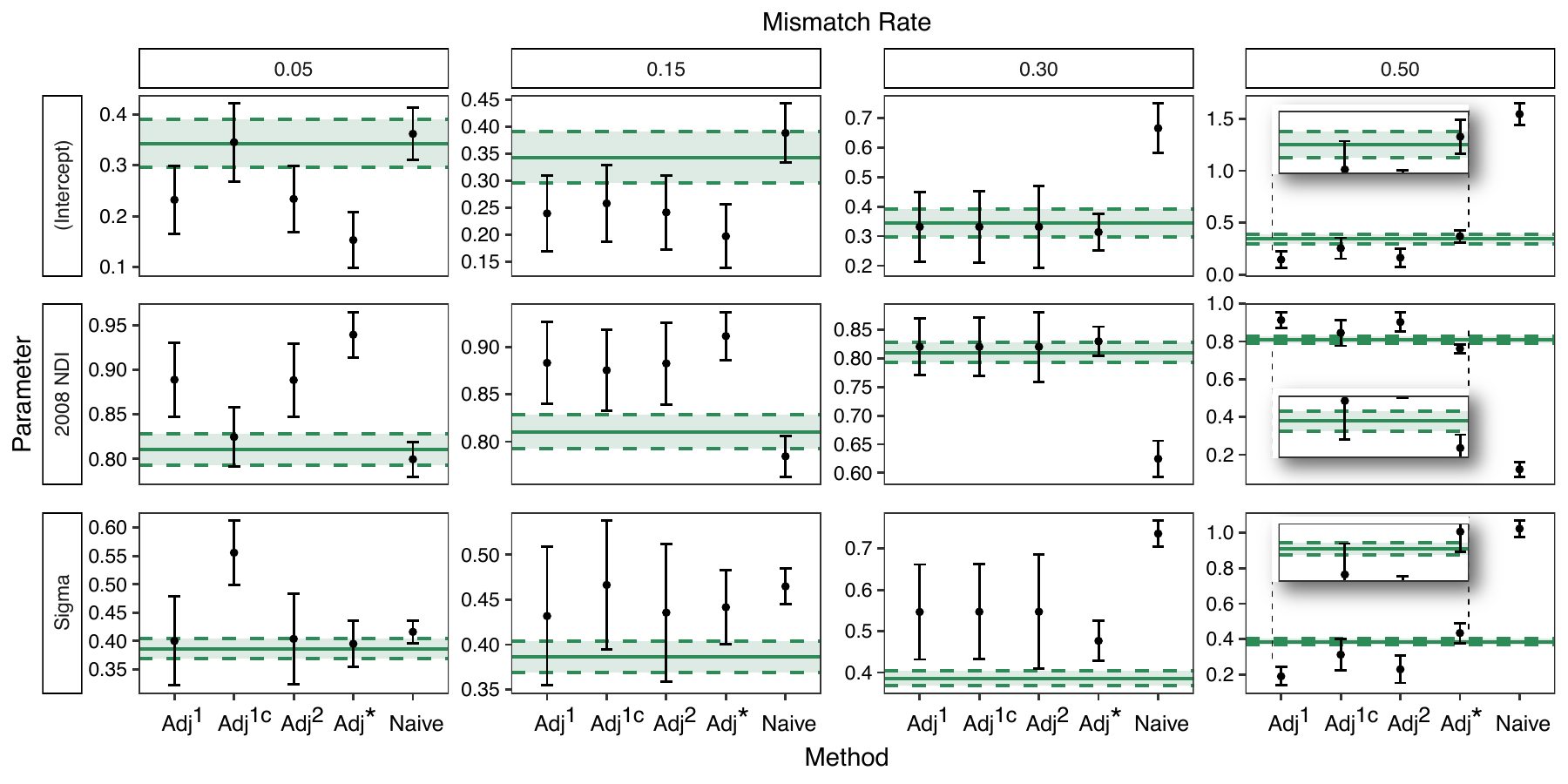}
\hspace*{-1ex}
\vspace{-5ex}
\caption{Under the different mismatch rate settings, 95\% confidence intervals for the parameters from the analyses. Estimates are plotted as points, with the error bars showing the lower and upper interval bounds. As a benchmark, the oracle results (i.e., ideal analysis of perfectly linked file) are shown in green, with estimates plotted using solid lines and the dashed lines representing 95\% confidence interval bounds. In the rightmost panel, we provide boxes that  ``zoom in" the ranges indicated by dashed lines.}\label{fig:cs_est}
\end{figure}

The coefficient estimates (and 95\% confidence intervals) from the varied analyses for the outcome model of interest are depicted in Figure \ref{fig:cs_est}. Results for the latent mismatch indicator model of secondary interest as well as standard errors are reported in Table \ref{tab_cs}. 

The Naive approach is significantly impacted with higher mismatch rates. Its estimates for the predictor-response relationship ($\beta_1$) and the residual standard deviation ($\sigma$) exhibit substantial attenuation towards the null, which is an expected consequence of ignoring mismatch error. While the Naive approach produces relatively precise (i.e., narrow) confidence intervals, these intervals can deviate greatly from the oracle results, rendering the approach unsuitable for reliable inference, particularly at mismatch rates higher than 15\%.

In contrast, the adjustment approaches yield results that are generally comparable to the oracle baseline. The proposed approach becomes increasingly useful as the mismatch rate increases. Compared to the plain adjustment approaches, the standard errors from the proposed approach are relatively more stable  across all mismatch rates and the confidence intervals tend to be narrower. While estimates from the adjustment approaches are often similar at moderate mismatch rates (15\% and 30\%), notable differences are evident for the parameter estimates at the 5\% and 50\% settings. At a low mismatch rate (5\%), violation of the strongly non-informative linkage assumption is not expected to be as severe. At the same time, we observe difficulty with modeling the latent mismatch indicator and over-adjustment. Based on the results from plain adjustment with versus without the constraint on the estimated overall mismatch rate (Adj$^1$ and Adj$^{1c}$, respectively), imposing a constraint on the assumed mismatch rate in the proposed approach may similarly facilitate modeling. On the other hand, at a high mismatch rate (50\%), the proposed approach shows minimal deviation from the oracle, highlighting its effectiveness in correcting for bias induced by the mismatches.

\begin{table}[]
\centering
{\small \begin{tabular}{|c|c|c|c|c|}
\hline
\multirow{2}{*}{Mismatch Rate} &        & $\widehat{\beta}_0$     & $\widehat{\beta}_1$     & $\widehat{\sigma}$      \\ \cline{2-5} 
                               & Oracle & 0.3429 (0.024) & 0.8104 (0.009) & 0.3862 (0.009)\\ \hline
\multirow{4}{*}{0.05}             & Naive  & 0.3614 (0.026) & 0.7997 (0.01) & 0.4161 (0.01) \\ \cline{2-5} 
                               & Adj$^1$   & 0.2320 (0.034) & 0.8889 (0.021) & 0.4001 (0.04) \\ \cline{2-5} 
                               & Adj$^{1c}$  & 0.3449 (0.039) & 0.8247 (0.017) & 0.5558 (0.029) \\ \cline{2-5} 
                               & Adj$^{2}$  & 0.2337 (0.033) & 0.8885 (0.021) & 0.4038 (0.041) \\ \cline{2-5}          
                               & Adj$^{*}$   & 0.1531 (0.028) & 0.9394 (0.013) & \textbf{0.3951} (0.021) \\ \hline
\multirow{4}{*}{0.15}            & Naive  & 0.3879 (0.028) & 0.7844 (0.011) & 0.4648 (0.010) \\ \cline{2-5} 
                               & Adj$^1$   & 0.2391 (0.036) & 0.8832 (0.022) & 0.4318 (0.039) \\ \cline{2-5} 
                               & Adj$^{1c}$  & 0.2576 (0.036) & 0.8754 (0.025) & 0.4664 (0.036) \\ \cline{2-5} 
                               & Adj$^{2}$  & 0.2410 (0.035) & 0.8826 (0.022) & 0.4356 (0.039) \\ \cline{2-5} 
                               & Adj$^{*}$   & 0.1969 (0.03) & 0.9115 (0.013) & 0.4416 (0.021) \\ \hline
\multirow{4}{*}{0.30}            & Naive  & 0.6646 (0.043) & 0.6247 (0.016) & 0.7361 (0.016) \\ \cline{2-5} 
                               & Adj$^1$   & 0.3312 (0.060) & 0.8204 (0.025) & 0.5465 (0.059) \\ \cline{2-5} 
                               & Adj$^{1c}$  & 0.3313 (0.061) & 0.8203 (0.026) & 0.5469 (0.059) \\ \cline{2-5} 
                               & Adj$^{2}$  & 0.3314 (0.071) & 0.8203 (0.031) & 0.5470 (0.07) \\ \cline{2-5} 
                               & Adj$^{*}$   & 0.3135 (0.031) & 0.8299 (0.013) & \textbf{0.4765} (0.025) \\ \hline
\multirow{4}{*}{0.50}            & Naive  & 1.5432 (0.053) & 0.1227 (0.02) & 1.0220 (0.023) \\ \cline{2-5} 
                               & Adj$^1$   & 0.1454 (0.041) & 0.9134 (0.022) & 0.1908 (0.026) \\ \cline{2-5} 
                               & Adj$^{1c}$  & 0.2550 (0.051) & 0.8459 (0.035) & 0.3128 (0.045) \\ \cline{2-5} 
                               & Adj$^{2}$  & 0.1647 (0.045) & 0.9032 (0.026) & 0.2302 (0.039) \\ \cline{2-5} 
                               & Adj$^{*}$   & \textbf{0.3708} (0.03) & 0.7624 (0.012) & \textbf{0.4345} (0.029) \\ \hline
\end{tabular}

\vspace{2ex}
\centering
\addtolength{\tabcolsep}{-1.5pt}
\begin{tabular}{|l|ll|ll|ll|ll|}
\hline
\multirow{2}{*}{} & \multicolumn{2}{l|}{0.05}       & \multicolumn{2}{l|}{0.15}        & \multicolumn{2}{l|}{0.30}         & \multicolumn{2}{l|}{0.50}          \\ \cline{2-9} 
                  & \multicolumn{1}{l|}{$\widehat{\gamma}_0$} & $\widehat{\gamma}_1$ & \multicolumn{1}{l|}{$\widehat{\gamma}_0$}   & $\widehat{\gamma}_1$ & \multicolumn{1}{l|}{$\widehat{\gamma}_0$}   & $\widehat{\gamma}_1$  & \multicolumn{1}{l|}{$\widehat{\gamma}_0$}   & $\widehat{\gamma}_1$   \\ \hline
True              & \multicolumn{1}{l|}{2}  & 3  & \multicolumn{1}{l|}{0.45} & 3  & \multicolumn{1}{l|}{0.45} & 0.1 & \multicolumn{1}{l|}{0.45} & -0.4 \\ \hline
Adj$^2$ &
  \multicolumn{1}{l|}{\begin{tabular}[c]{@{}l@{}}1.3286\\ (0.192)\end{tabular}} &
  \begin{tabular}[c]{@{}l@{}}-0.0087\\ (0.005)\end{tabular} &
  \multicolumn{1}{l|}{\begin{tabular}[c]{@{}l@{}}1.3201\\ (0.185)\end{tabular}} &
  \begin{tabular}[c]{@{}l@{}}-0.0084\\ (0.005)\end{tabular} &
  \multicolumn{1}{l|}{\begin{tabular}[c]{@{}l@{}}0.6658\\ (0.175)\end{tabular}} &
  \begin{tabular}[c]{@{}l@{}}-0.0007\\ (0.004)\end{tabular} &
  \multicolumn{1}{l|}{\begin{tabular}[c]{@{}l@{}}-0.1801\\ (0.178)\end{tabular}} &
  \begin{tabular}[c]{@{}l@{}}-0.4427\\ (0.112)\end{tabular} \\ \hline
Adj$^*$ &
  \multicolumn{1}{l|}{\begin{tabular}[c]{@{}l@{}}1.1070\\ (0.137)\end{tabular}} &
  \begin{tabular}[c]{@{}l@{}}-0.0144\\ (0.008)\end{tabular} &
  \multicolumn{1}{l|}{\begin{tabular}[c]{@{}l@{}}1.1040\\ (0.138)\end{tabular}} &
  \begin{tabular}[c]{@{}l@{}}-0.0044\\ (0.004)\end{tabular} &
  \multicolumn{1}{l|}{\begin{tabular}[c]{@{}l@{}}0.1907\\ (0.121)\end{tabular}} &
  \begin{tabular}[c]{@{}l@{}}-0.0004\\ (0.003)\end{tabular} &
  \multicolumn{1}{l|}{\begin{tabular}[c]{@{}l@{}}0.2392\\ (0.184)\end{tabular}} &
  \begin{tabular}[c]{@{}l@{}}-0.5275\\ (0.147)\end{tabular} \\ \hline
\end{tabular}}
\caption{Coefficient estimates (with standard errors in parentheses) for the case study. Outcome model estimates from the proposed approach that are relatively close to the oracle approach and better than those of the other approaches are indicated in bold.}\label{tab_cs}
\end{table}

\spacingset{1.45}
\section{Conclusion}\label{sec:conclusion}
\vspace*{-1.5ex}
In this paper, the general framework in \cite{slawski2024general} is extended to enable linkage to depend on covariates of primary interest in the linked data for analysis. This assumption often does not hold in practice and can be especially challenging to address in the secondary setting with existing approaches. Empirical investigation of the approach shows that the extension can be useful for post-linkage data analysis. At the same time, the present work leads to several avenues for future research. Based on the case study, one direction is to develop an approach to incorporate prior information on the underlying mismatch rate if available. This may help with improving estimation of the parameters for the unobserved mismatch indicators. Another important direction is enhancing the scalability of the proposed approach. Further evaluation of the proposed method under model misspecification is also needed. Finally, although empirical evaluation here is focused on linear regression, the proposed approach is applicable to other regression settings.

\newpage
\bibliographystyle{abbrvnat}
\renewcommand\bibpreamble{\vspace{-0.15\baselineskip}} 
\setlength{\bibsep}{1.2ex}
{\small \bibliography{references_M_short}}
\clearpage
\appendix
\noindent 
{\bfseries {\large Appendix}}

\setcounter{page}{1}
\subsubsection*{Proof of Lemma 1}

The marginal density of $\M{y}|\{ m = 1\}$ takes the form
\begin{align*}
f(\M{y} | m = 1) &= \int f(\M{y} | m = 1, \M{x}, \M{z}) \, f(\M{x}, \M{z} | m = 1) \; dP(\M{x}, \M{z}) \\
                 &=  \int f(\M{y} | \M{x}) \, \dfrac{\M{P}(m = 1 | \M{x}, \M{z})}{\int \M{P}(m = 1 | \M{x}', \M{z}') \, dP(\M{x}', \M{z}')}  \, dP(\M{x}, \M{z}) \\
                 &=\int f(\M{y} | \M{x}) \, \dfrac{\M{P}(m = 1 | \M{z})}{\int \M{P}(m = 1 | \M{z}') \, dP(\M{x}, \M{z}')}  \, dP(\M{x}, \M{z}) \\
                 &=  \sum_{j = 1}^n f(\M{y}|\M{x}_j;\bm{\theta}) \, \frac{1 - h(\M{z}_j;\bm{\gamma})}{\sum_{k = 1}^n (1 - h(\M{z}_k;\bm{\gamma}))} = \sum_{j = 1}^n f(\M{y}|\M{x}_j;\bm{\theta}) \; \omega_j(\M{z}_j;\bm{\gamma}),
\end{align*}
where $P$ denotes the probability measure placing mass $1/n$ at each of the $\{ (\M{x}_i, \M{z}_i)  \}_{i=1}^n$. Note that in the second inequality we use that $\M{y} \indep \{m, \M{z}\} | \M{x}$, and the third equality, we use that $m \indep \M{x} | \M{z}$ (cf.~Figure \ref{fig:dag}). In the last equality, we simplify the integral against the measure $P$ to an average over $\{ (\M{x}_i, \M{z}_i)  \}_{i=1}^n$.  



\subsubsection*{Dependence of $m$ on both $\M{x}$ and $\M{y}$}

\noindent{}Suppose that $m$ depends on both $\M{y}$ and $\M{x}$. From Bayes' formula, we have 
\begin{equation*}
    f( \M{y}| \M{x}, m = 0) = \dfrac{\mathbf{P}(m=0| \M{x},  \M{y}) \cdot f( \M{y}| \M{x})}{\mathbf{P}(m=0| \M{x})}.
\end{equation*}

\noindent{}In general, only if $ \M{y} \indep m| \M{x}$, we have $f( \M{y}| \M{x},m = 0) = f( \M{y}| \M{x})$. If the latter relationship does not hold, the relationship between $ \M{x}$ and $ \M{y}$ for correct matches is different from the ones assumed for the population. In particular, the $ \M{y} |  \M{x}$ model in the sub-population of correct matches may no longer be equal to a linear regression model if originally assumed. While the relationship $\M{y}|\M{x}, m = 0$ may still be estimable, inferring the relationship $y | \M{x}, m = 1$ appears to be impossible even if the mismatch indicators 
are known -- unless additional assumptions are made. Indeed, for mismatches, we do not 
observe correctly paired $(\M{x}, \M{y})$ that would allow us to do so.

\end{document}